\documentclass{aa}
\usepackage{graphicx} 
\usepackage[varg]{txfonts}
\usepackage{natbib,twoopt}
\usepackage{adjustbox}
\usepackage{placeins}
\usepackage{longtable} 
\usepackage{lscape}    
\usepackage[breaklinks=true]{hyperref} 
\bibpunct{(}{)}{;}{a}{}{,}             
\makeatletter
  \newcommandtwoopt{\citeads}[3][][]{\href{http://adsabs.harvard.edu/abs/#3}%
    {\def\hyper@linkstart##1##2{}%
     \let\hyper@linkend\@empty\citealp[#1][#2]{#3}}}
  \newcommandtwoopt{\citepads}[3][][]{\href{http://adsabs.harvard.edu/abs/#3}%
    {\def\hyper@linkstart##1##2{}%
     \let\hyper@linkend\@empty\citep[#1][#2]{#3}}}
  \newcommandtwoopt{\citetads}[3][][]{\href{http://adsabs.harvard.edu/abs/#3}%
    {\def\hyper@linkstart##1##2{}%
     \let\hyper@linkend\@empty\citet[#1][#2]{#3}}}
  \newcommandtwoopt{\citeyearads}[3][][]%
    {\href{http://adsabs.harvard.edu/abs/#3}
    {\def\hyper@linkstart##1##2{}%
     \let\hyper@linkend\@empty\citeyear[#1][#2]{#3}}}
\makeatother

\title{A multi-frequency{\bf ,} multi-epoch radio continuum study of the Quintuplet cluster with the Very Large Array}
\author{
        M. Cano-Gonz\'alez
         \inst{1}
          \and
        R. Sch\"odel
         \inst{1}
          \and
          A. Alberdi
         \inst{1}
         \and
         J. Moldón
         \inst{1}
          \and
          M. P\'erez-Torres
         \inst{1,2}
         \and
          F. Najarro
         \inst{3}
         \and
         A.T. Gallego-Calvente
         \inst{1,4}
}
\institute{
           Instituto de Astrof\'isica de Andaluc\'ia (CSIC),
           Glorieta de la Astronom\'ia s/n, 18008 Granada, Spain.
           \email{mcano@iaa.es}
           \and
           School of Sciences, European University Cyprus, Diogenes street, Engomi, 1516 Nicosia, Cyprus
           \and
           Centro de Astrobiolog\'ia, CSIC-INTA, Ctra de Torrej\'on a Ajalvir km 4, 28850 Torrej\'on de Ardoz, Madrid, Spain
           \and 
           Universitat de València. Departament d'Astronomia i Astrofísica. Facultat de Física. Av. Doctor Moliner, 50 - Burjassot, Valencia 
}
\date{}

\begin{document}
\abstract
{The  Quintuplet cluster, located in the Galactic Centre, is one of the few young massive clusters in the Milky Way. It allows us to study dozens of massive, post main-sequence stars individually, providing unique insights into the properties of the most massive stars.} 
{Our goal is to study the radio continuum emission of  the most massive stars in the cluster.}
{We carried out a total of nine observations (three in the  C-band and six in the X-band) of the Quintuplet cluster with the Karl G. Jansky Very Large Array in the A-configuration. We cross-matched the detected sources with infrared stellar catalogues to ensure cluster membership, calculated their spectral indices, quantified variability, and inferred clumping-scaled mass-loss rates.}
{We present the most complete catalogue of radio stars in the Quintuplet cluster to date, with a total of 41 detections, and the deepest images of the cluster in the 4 to 12 GHz range (reaching an rms noise level of $2.3\, \mu\mathrm{Jy/beam}$ in the X-band). The six year baseline of our observations allowed us to perform a robust variability assessment, finding that around $60\%$ of the Quintuplet radio stars are variable on timescales of months to years. We derived the spectral indices of 28 out of the 41 sources. Based on their spectral indices and variability, we classified 11 of them as colliding-wind binaries, seven as strictly thermal sources, and ten as ambiguous. Including the ambiguous sources, we estimate a multiplicity fraction of ($75\pm22\%$). We also computed upper limits for the mass-loss rates of the thermal radio stars, finding them in agreement with typical values for Wolf-Rayet (WNh and WC) stars. Finally, we compared these results to the ones obtained from our analogous study of the Arches cluster.} 
{}

\maketitle

\section{Introduction}
Massive  O-type stars  form in small numbers and have short lives  ($<10$\,Myr on the main sequence). They are therefore rare and typically located at large distances. Most massive stars form in binaries or higher-order multiples, with their interaction frequently affecting their evolution through different mechanisms, such as mass transfer, common envelopes, and rejuvenation  \citep{Sana2012,Marchant_Bodensteiner2024}.  For these reasons it is challenging to gain a complete understanding of the phases that massive stars go through after they leave the main sequence. 

Radio continuum observations of massive stars can inform us about their mass-loss and multiplicity through two main, non-exclusive emission mechanisms. On the one hand,  thermal (free-free) emission  arises from  ionised line-driven winds  \citep{Castor_Abbot_Klein1975} and can allow mass-loss rates to be estimated.  Optically thick winds, such as those present in Wolf-Rayet stars, are particularly bright in the radio spectrum. On the other hand, in the case of a binary (or multiple) system, non-thermal (synchrotron) emission is emitted from the  colliding-wind region between two  stars \citep{DeBecker2007,DeBecker2013} if both of them have sufficiently strong  winds. Thus, in a colliding-wind binary (CWB), the observed radio emission is a combination of the thermal component arising from the winds of each member and the non-thermal contribution formed in the colliding wind region between the stars \citep{DeBecker2013}.

The orbital parameters of binaries determine the strength and detectability of the non-thermal component of their radio emission. In a system with a short orbital period (less than two weeks, approximately), the non-thermal emission may be trapped within the optically thick area of the combined ionised wind region, thus masking the observational traces of binarity. In systems with eccentric orbits  the dominant emission mechanism may therefore  vary as a function of orbital phase, as the non-thermal component can become absorbed by the thermal wind envelope when the stars approach each other around periastron \citep{Dougherty2005, Sanchez-Bermudez2019}.

Assuming that the flux density ($S_\nu$) scales with observing frequency ($\nu$) as $S_\nu\propto \nu^\alpha$, the spectral response is given by the spectral index, $\alpha$. Thermal emission is characterised by a positive `canonical' value of $\alpha\approx0.6$, assuming a spherically symmetric, fully ionised wind \citep{Wright_Barlow1975, Panagia_Felli1975}. In contrast, non-thermal emission is characterised by lower values of the spectral index ($\alpha\lesssim 0.0$). Positive $\alpha$ values that are lower than the canonical value of $\alpha\approx0.6$ may indicate the presence of a non-thermal component  \citep{DeBecker2007}.

Different observational techniques are sensitive to different ranges of orbital separations in multiple systems \citep[see  Fig.\,1 in][]{Sana2011}. Radial velocity spectroscopic studies typically cover separations up to 10\,AU and periods up to a thousand days,  interferometric techniques cover separations of a few to a few tens of AU and periods from a few hundred days to a few decades, and high angular resolution imaging techniques cover separations upwards from a few tens of AU and periods  longer than a hundred years. X-ray and radio continuum observations are  sensitive to systems with periods that may range from a few weeks to years \citep[e.g.][]{Sanchez-Bermudez2019,Dougherty2005}. For example, a recent X-ray study of Westerlund 1 shows that most of the Wolf-Rayet population of the cluster ($\sim90\%$) present observational traces of binarity in the form of unresolved X-ray emission associated with a massive star \citep{Anastasopoulou2024}.  Radio observations of the Arches cluster indicate a multiplicity fraction greater than $60\%$ \citep{Arches_paper}. Multi-epoch, near-infrared, radial velocity studies of the Arches cluster have also revealed high multiplicity fraction, greater than $60\%$ \citep{Clark_IV}.

Given the rareness of massive stars, young massive clusters are ideal laboratories to study several dozens to hundreds of massive stars with the same age and metallicity, located at the same distance. The Galactic Centre \citep[GC; at 8.2 kpc from Earth,][]{GRAVITY2019} harbours three young massive clusters: the massive stars within 0.5\,pc of  the massive black hole Sgr A* \citep[e.g.][]{Paumard2006,Do2013} and the Arches and Quintuplet clusters, both of which are around 30 pc in projection from Sgr A* \citep{Figer1999,Clark_I,Clark2018_Quintuplet}. The Arches and the Quintuplet clusters have masses of $1-2\times10^4\, M_\odot$, but Arches is about one million years younger than Quintuplet \citep{Liermann2012,Clark_IV}. The Arches and Quintuplet clusters have both been targeted previously by radio observations \citep[e.g.][]{Lang2005,G-C2022,G-C2021}.

Recently, we have presented a new analysis of the radio stars in the Arches cluster, based on the deepest, longest time baseline Karl. G. Jansky Very Large Array (VLA) observations. These observations allowed us to report the greatest number of radio-detected stars in this cluster as well as a robust determination of their mass-loss rates, variability, and multiplicity \citep{Arches_paper}. Since there are massive stars scattered throughout the GC environment, that may not necessarily be directly related to one of the massive clusters, we also made use of published  infrared proper motion measurements to assess the cluster membership of the  stars. In this paper, we present an equivalent radio continuum study of the Quintuplet cluster.

This work is organised as follows. In Sect. \ref{section_obs} we describe the observations and present the imaging process. Section \ref{section_analysis} presents the flux extraction, variability assessment, and spectral index derivation processes and shows the main results of the study. In Sect. \ref{sect_discussion_conclusion} we discuss the main results, and finally, Sect. \ref{sect_conclusions} shows the main conclusions.

\section{Observations and imaging}\label{section_obs}
We carried out a total of nine observations of the Quintuplet cluster ($\alpha^\mathrm{J2000}=17^{\mathrm{h}}\, 46^{\mathrm{m}}\, 15^{\mathrm{s}}$, $\delta^\mathrm{J2000}=-28\degr\, 49\arcmin\, 33\arcsec$) with the VLA across three different years: 2016, 2018, and 2022. We used C- and X-bands (central frequencies of 6 and 10 GHz respectively) in the A-configuration to characterise the emission caused by the strong winds of individual cluster members. We show details of the observations and their related image properties in Table \ref{table_observations}.

We used the VLA pipeline included in the Common Astronomy Software Applications (CASA; \citealt{CASA}) version 6.4.1, to perform standard radio continuum calibration. We note that the data from 2016 and 2018 epochs are the same as those of \citet{G-C2022}, but we re-calibrated them from scratch. For all observations, we used 3C286 (J1331+3030) as the flux density and bandpass calibrator. Regarding phase calibration, J1744--3116 was used in 2016 and 2018, in contrast to 2022 observations which used J1820--2528. The main reason for the change in phase calibrator was that J1744--3116 presented some extended emission from its jet. However, we found no systematic differences (concerning flux extraction and image quality) between reducing the data with either phase calibrator. 

We used the {\tt tclean} task to deconvolve the dirty images. This task uses different modern versions of the CLEAN algorithm \citep{Hogbom74} within a Python interface. We used the multi-term, multi-frequency synthesis algorithm ({\tt deconvolver=mtmfs}, \citealt{Rau_&_Cornwell2011}) and {\tt gridder=widefield}, as well as 4 to 5 pixels across the synthesised beam ($\approx0\farcs33$ for C-band and $\approx0\farcs2$ for X-band in A-configuration) which resulted in pixel sizes of $0\farcs067$ and $0\farcs05$ for the C- and X-bands respectively. As in \citet{G-C2022}, we set the {\tt gain} parameter to 0.05 in order to spot sources during the interactive cleaning. We tried different weightings. We obtained the best compromise between off-source rms noise, point source characterisation, and resolution with Briggs weighting ({\tt robust=0.5}). 

The Quintuplet field lies within the so-called radio-bright zone of the GC, and, in particular, it is seriously affected by the extended emission of non-thermal radio filaments (\citealp[NTFs; e.g.][]{Pare2019,Heywood2022}). Therefore, while creating the images, we used a $>200\mathrm{k}\lambda$ $u$-$v$ cut for both bands, excluding short baselines, which allowed us to minimise the contribution of the extended emission. However, even after applying the \textit{u-v} cut, remnants of the extended emission were still present and over-resolved by the VLA in A-configuration; especially bright areas, such as the pistol nebula, suffer from this effect more prominently. Using the multiscale algorithm did not improve image quality because it included negative, spurious unphysical values into the cleaned models. Thus, we discarded the multiscale approach and interactively masked point sources only. Finally, we imposed a stopping criterion of \texttt{nsigma=5} for all of our images.

We self-calibrated each dataset in both phase and amplitude. To do so, we combined both polarisation channels while creating the calibration tables with {\tt gaincal} in order to obtain a higher S/N. The best solution intervals fell within the 5 to 10 s range (fewer than $5\%$ of the solutions have S/N$<5$). 

The presence of the bright ($\sim 44$ mJy and $\sim42$ at C- and X-bands,
respectively) source N3 \citep{Ludovici2016} a few arcminutes away from the Quintuplet cluster severely contaminated preliminary images and intermediate products, making it necessary to iterate the self-calibration process. Most of our final images were produced after an initial phase-only self-calibration followed by a self-calibration in the \texttt{'ap'} mode, that is, in both phase and amplitude. Although a few artefacts surrounding N3 remained, the iterative self-calibration reduced them enough so that the emission from the cluster members was not affected. We note that we used the entire primary beam of the VLA ($\theta_{\mathrm{PB}}^{\mathrm{FWHM}}\approx7\arcmin$ and $\theta_{\mathrm{PB}}^{\mathrm{FWHM}}\approx4.2\arcmin$ for C- and X-band respectively) during the self-calibration and imaging process. We added to the cleaned model those point sources present in the field of view that presumably do not belong to the Quintuplet cluster (such as N3) in order to populate it with the most real emission possible. An analysis of these field sources is left for future work.

After self-calibration, we corrected for primary beam attenuation with the CASA task {\tt impbcor} and created a reduced image centred on the cluster ($\approx2\arcmin\times2\arcmin$) large enough to encompass all NIR sources from the massive stellar catalogue by \citet{Clark2018_Quintuplet}. We used the task {\tt concat} to concatenate measurement sets from different observations of a given epoch and created the combined images of 2016 and 2022. We also used the same task to create deep images, in which all visibilities for a given band are concatenated (see Table \ref{table_observations} and Fig. \ref{Quint_deep_X}). In order to account for systematic positional offsets between epochs when creating the deep images, it was necessary to globally self-calibrate in phases (setting \texttt{solint='inf'} while creating the self-calibrating tables with {\tt gaincal}).

\section{Analysis and results}\label{section_analysis}

\subsection{Flux extraction and astrometry}

We extracted point source flux densities using the CASA task \texttt{imfit}. This task fits a single Gaussian to a region given by the user and returns peak flux, integrated flux density and the fit's coordinates along with their related uncertainties following the prescription by \citet{Condon1997}. We used circular regions with $1\farcs5$ and $2\arcsec$ diameter for X- and C-bands respectively. This way, we gathered enough surface to compute a reliable local rms for each fit. Increasing the region area up to $3\arcsec$ did not improve the extracted fluxes and their uncertainties in a significant manner; in fact it worsened uncertainties artificially because of close-neighbour contamination in some instances. In addition, we took into account systematic flux errors due to absolute calibration. We computed them by assuming a conservative $5\%$ and $3\%$ of a particular detection's flux density for C- and X-bands, respectively \citep{Perley_Butler2017} and added them quadratically. Using our X-band deep image, we detected a total of 41 radio point sources within the inner $\approx 2\arcmin\times2\arcmin$ of the Quintuplet cluster. Thus, the deep X-band image is sensitive to radio stars with flux densities $\gtrsim12\, \mu\mathrm{Jy}$.

Coordinate uncertainties returned by the {\tt imfit} task are partly dependent on the S/N of the point source. Therefore, we referred all coordinates extracted from the individual observations to the X-band deep image, as it shows the best off-source rms. Then, we used \citet{Hosek2022} catalogue as the astrometric reference frame. We selected their sources with a cluster membership probability greater than $0.7$ and cross-matched them with our deep X-band radio point sources. After the initial offset correction, our detections showed a mean coordinate offset of $-0\farcs0003 \pm 0\farcs007$ in right ascension and
$0\farcs002 \pm 0\farcs011$ in declination. The errors represent the standard deviation of the final offsets. Thus, we computed the total coordinate uncertainties as

\begin{equation}
    \sigma_{\mathrm{T}} = \sqrt{\sigma_{\mathrm{fit}}^2 + \sigma_{\mathrm{std}}^2 + \sigma_{\mathrm{\theta}}^2}
\end{equation}
where $\sigma_{\mathrm{fit}}$ is the error returned by {\tt imfit}, $\sigma_{\mathrm{std}}$ is the standard deviation of the aforementioned offset with respect to the \citet{Hosek2022} catalogue, and $\sigma_{\theta}$ is another quantity that takes into account channel width and distance to phase centre in the form $\sigma_{\mathrm{\theta}}=r_\theta\times\frac{\nu_c}{\nu_0}$ ($\nu_c=2000\,\mathrm{kHz}$, $\nu_0=10\,\mathrm{GHz}$, \citealt{Thompson2017}). Table \ref{fluxtable} shows the final coordinates of our radio sources along with their flux densities and related uncertainties.

\subsection{Cross-matching with infrared stellar catalogues}\label{subsect_ID}
\begin{figure*}
   \sidecaption
   \includegraphics[width=12cm]{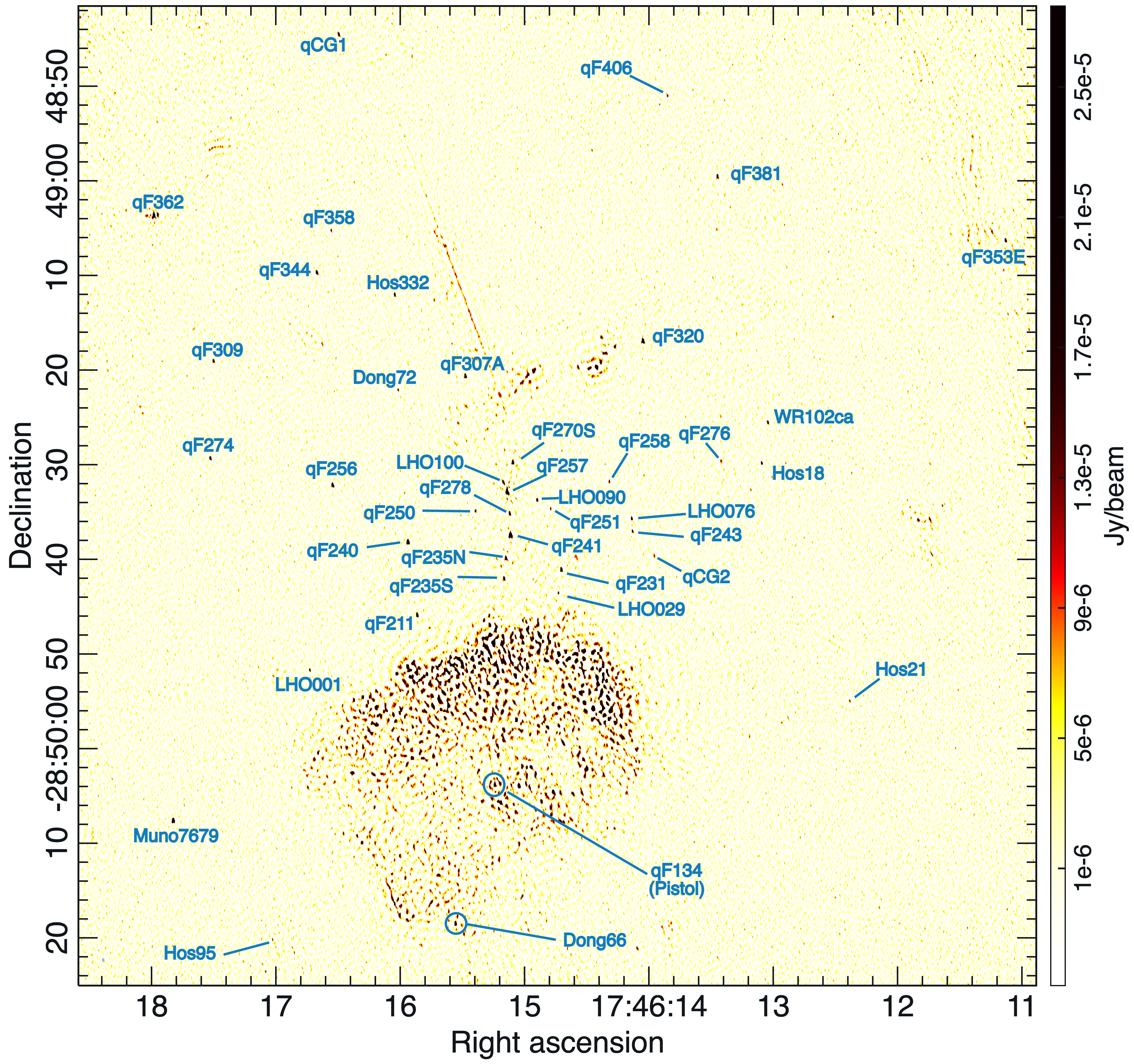}
      \caption{Deep X-band image of the Quintuplet cluster (see Table \ref{table_observations} for details). Sources are labelled with the NIR stellar ID from \citet{Clark_I}, \citet{Dong2011}, \citet{Muno2009} or \citet{Hosek2022} (see Sect. \ref{subsect_ID}).}
         \label{Quint_deep_X}
   \end{figure*}

In order to identify the radio point sources, we cross-matched them with the \citet{Clark2018_Quintuplet} massive stellar catalogue of the Quintuplet cluster, \citet{Dong2011} \textit{Hubble Space Telescope} Paschen-$\alpha$ emitters, \citet{Muno2009} \textit{Chandra} X-ray point sources, and the \citet{Hosek2022} proper motion catalogue. Table \ref{fluxtable} shows the stellar ID from \citet{Clark2018_Quintuplet}, \citet{Dong2011}, \citet{Muno2009}, or \citet{Hosek2022}. Figure \ref{Quint_deep_X} shows the positions of the radio stars with their respective ID in the X-band deep image.

\subsection{Variability assessment}\label{subsect_variability}
We used a similar procedure to that of \citet{Zhao2020} to compute radio variability. Namely, for a given band, we imposed $\Delta S/\sigma>5$ for a particular point source to be noted as variable where $\Delta S=S_\nu^{\mathrm{max}}-S_\nu^{\mathrm{min}}$ and

\begin{equation}
    \sigma = \left( \sum_{i=1}^N \frac{1}{(\sigma_{S_{\nu,i}})^2}  \right)^{-1/2}.
\end{equation}
In the previous expressions, $S_\nu^{\mathrm{max(min)}}$ represents the maximum (minimum) flux density observed (across all observations) for a particular source at a given frequency, $\nu$; $N$ is the number of observations in which a source is detected in a given band; and $\sigma_{S_{\nu,i}}$ is the flux density uncertainty of a particular source in the $i$-th observation. Thus, we used three and six observations to compute variability for the C- and X-bands, respectively. For those faint sources that only appeared as $5\sigma$ detections in the concatenated data of 2016 or 2022, we followed an analogous procedure with the combined images. We note that for this variability analysis, we did not take into account the $5\%$ and $3\%$ systematic flux errors due to absolute calibration because they affected all sources equally. In all, we computed variability for 36/41 point sources, with 22 of them showing variability on the $5\sigma$  level.

\subsection{Spectral indices}
We followed the same procedure as in \citet{Arches_paper} to derive the spectral indices of the Quintuplet radio detections. Namely, we followed two methods: First, we created four sub-band images of 1 GHz bandwidth for each individual observation with the \texttt{tclean} task. Then, for each source detected in all sub-band images of a particular observation (minimum of four points to be fitted), we performed a linear least-squares fit with inverse variance weighting in flux versus frequency logarithmic space. Thus, the slope of the fit returns the spectral index of a particular source. Moreover, since in 2022 we have C- and X-band observations carried out one day apart, we concatenated these measurement sets and obtained a dataset ranging a wider bandwidth (from 4 to 12 GHz) from which eight sub-band images were obtained, and therefore eight points to be fitted. The spectral index values from these combined C- and X-band data are shown in the second-to-last column of Table \ref{Table_alphas}. 

Secondly, for those sources not bright enough to be detected in all sub-band images but that are detected in both bands, we computed the spectral indices with their corresponding flux densities at the C- and X-bands as in \citet[][their Eqs. 3 and 4]{Arches_paper}. Table \ref{Table_alphas} shows the spectral indices of the Quintuplet radio stars.

Moreover, we also took into account the possibility of spectral index variations throughout the observations. Again, we followed the same procedure as in \citet{Arches_paper}, namely,
\begin{equation}\label{eq_alphavar}
    \Xi = \frac{\mathrm{max}(\alpha) - \mathrm{min}(\alpha)}{\left(\sum_i^M 1/(\sigma_{\alpha_i})^2  \right)^{-1/2}}
\end{equation}
where $M$ represents the number of spectral index values for a given source and $\sigma_{\alpha_i}$ is the spectral index uncertainty of a source in the $i$-th observation. Further discussion on these results is shown in Sect. \ref{subsect_alphavar}. 

\subsection{Proper motions and cluster membership}
We used the \citet{Hosek2022} proper motions catalogue of the Quintuplet cluster (bulk proper motions of $\mu_\alpha\, \cos\delta=-0.96\pm0.032$ and $\mu_\delta=-2.29\pm0.023\, \mathrm{mas\, yr^{-1}}$) to ensure cluster membership of the detected sources. Table \ref{Table_pm} shows their proper motion values for our detections. Most of Quintuplet's radio stars show high a cluster membership probability ($p_\mathrm{cluster}\gtrsim0.7$).
\begin{table}
\caption{Proper motions and cluster membership probability of the Quintuplet sources}
\label{Table_pm}
\centering
\begin{tabular}{l c c c}
\hline \hline
ID  & $\mu_\alpha\, \cos\delta$ &  $\mu_\delta$ & $p_\mathrm{cluster}$ \\
\hline
  qF257 & $-1.12 \pm 0.08$ & $-2.52 \pm 0.09$ & 0.64 \\ 
  qF241 & $-0.98 \pm 0.09$ & $-2.36 \pm 0.09$ & 0.93 \\ 
  qF240 & $-0.84 \pm 0.07$ & $-2.37 \pm 0.09$ & 0.90 \\ 
  qF320 & $-1.13 \pm 0.10$ & $-2.39 \pm 0.09$ & 0.84 \\ 
Muno7679 & $-0.82 \pm 0.16$ & $-2.50 \pm 0.18$ & 0.78 \\ 
 Dong66 & $-0.93 \pm 0.06$ & $-2.61 \pm 0.08$ & 0.48 \\ 
  qF256 & $-0.74 \pm 0.10$ & $-2.48 \pm 0.12$ & 0.70 \\ 
 qF270S & $-1.21 \pm 0.09$ & $-2.41 \pm 0.09$ & 0.63 \\ 
 qF235S & $-0.88 \pm 0.07$ & $-2.24 \pm 0.08$ & 0.94 \\ 
 qF307A & $-0.99 \pm 0.12$ & $-2.46 \pm 0.12$ & 0.86 \\ 
  qF344 & $-0.85 \pm 0.15$ & $-2.47 \pm 0.17$ & 0.82 \\ 
 qF235N & $-0.93 \pm 0.07$ & $-2.27 \pm 0.08$ & 0.94 \\ 
 qF353E & $-0.96 \pm 0.10$ & $-2.43 \pm 0.10$ & 0.89 \\ 
  qF381 & $-1.07 \pm 0.13$ & $-2.55 \pm 0.14$ & 0.71 \\ 
  qF278 & $-1.05 \pm 0.08$ & $-2.30 \pm 0.09$ & 0.92 \\ 
  qF309 & $-0.82 \pm 0.16$ & $-2.34 \pm 0.17$ & 0.86 \\ 
    LHO100 & $-0.96 \pm 0.09$ & $-2.48 \pm 0.09$ & 0.85 \\ 
  qF274 & $-1.04 \pm 0.14$ & $-2.42 \pm 0.15$ & 0.86 \\ 
 Hos332 & $-1.14 \pm 0.13$ & $-2.29 \pm 0.14$ & 0.84 \\ 
    LHO076 & $-1.02 \pm 0.07$ & $-2.27 \pm 0.07$ & 0.94 \\ 
WR102ca & $-1.14 \pm 0.07$ & $-2.20 \pm 0.06$ & 0.84 \\ 
    LHO090 & $-1.09 \pm 0.08$ & $-2.20 \pm 0.08$ & 0.89 \\ 
  qF243 & $-1.24 \pm 0.08$ & $-2.45 \pm 0.07$ & 0.46 \\ 
  Hos18 & $-1.11 \pm 0.08$ & $-2.32 \pm 0.07$ & 0.89 \\ 
  qF250 & $-0.75 \pm 0.30$ & $-2.66 \pm 0.33$ & 0.56 \\ 
    LHO001 & $-0.90 \pm 0.08$ & $-2.44 \pm 0.11$ & 0.89 \\ 
  qF258 & $-1.01 \pm 0.08$ & $-2.40 \pm 0.07$ & 0.92 \\ 
  qF358 & $-0.91 \pm 0.17$ & $-2.58 \pm 0.18$ & 0.71 \\ 
  qF251 & $-1.20 \pm 0.08$ & $-2.40 \pm 0.08$ & 0.70 \\ 
  Dong72 & $-0.88 \pm 0.10$ & $-2.50 \pm 0.12$ & 0.82 \\ 
  Hos21 & $-0.98 \pm 0.07$ & $-2.35 \pm 0.06$ & 0.94 \\ 
  Hos95 & $-0.73 \pm 0.13$ & $-2.64 \pm 0.15$ & 0.41 \\ 
  qF406 & $-1.47 \pm 0.16$ & $-2.32 \pm 0.17$ & 0.18 \\ 
  qF276 & $-0.88 \pm 0.08$ & $-2.56 \pm 0.07$ & 0.61 \\ 
    LHO029 & $-0.97 \pm 0.08$ & $-2.39 \pm 0.08$ & 0.92 \\ 
    qG29\tablefootmark{(a)} & $-2.30\pm0.11$ & $-3.59\pm0.08$ & 0.00 \\
\hline \hline
\end{tabular}
\tablefoot{Proper motion and cluster membership probability values from \citet{Hosek2022}. All proper motion units in milli-arcseconds per year $(\mathrm{mas}\,\mathrm{yr}^{-1})$. The counterparts of qF362, qF211, qF231, qCG1, qCG2, and qF134 were not found in \citet{Hosek2022} catalogue within $0\farcs2$ of our point sources. 
\tablefoottext{a}{ID from \citet{G-C2022}.}
}
\end{table}

\begin{figure}
  \centering
  \includegraphics[width=\hsize]{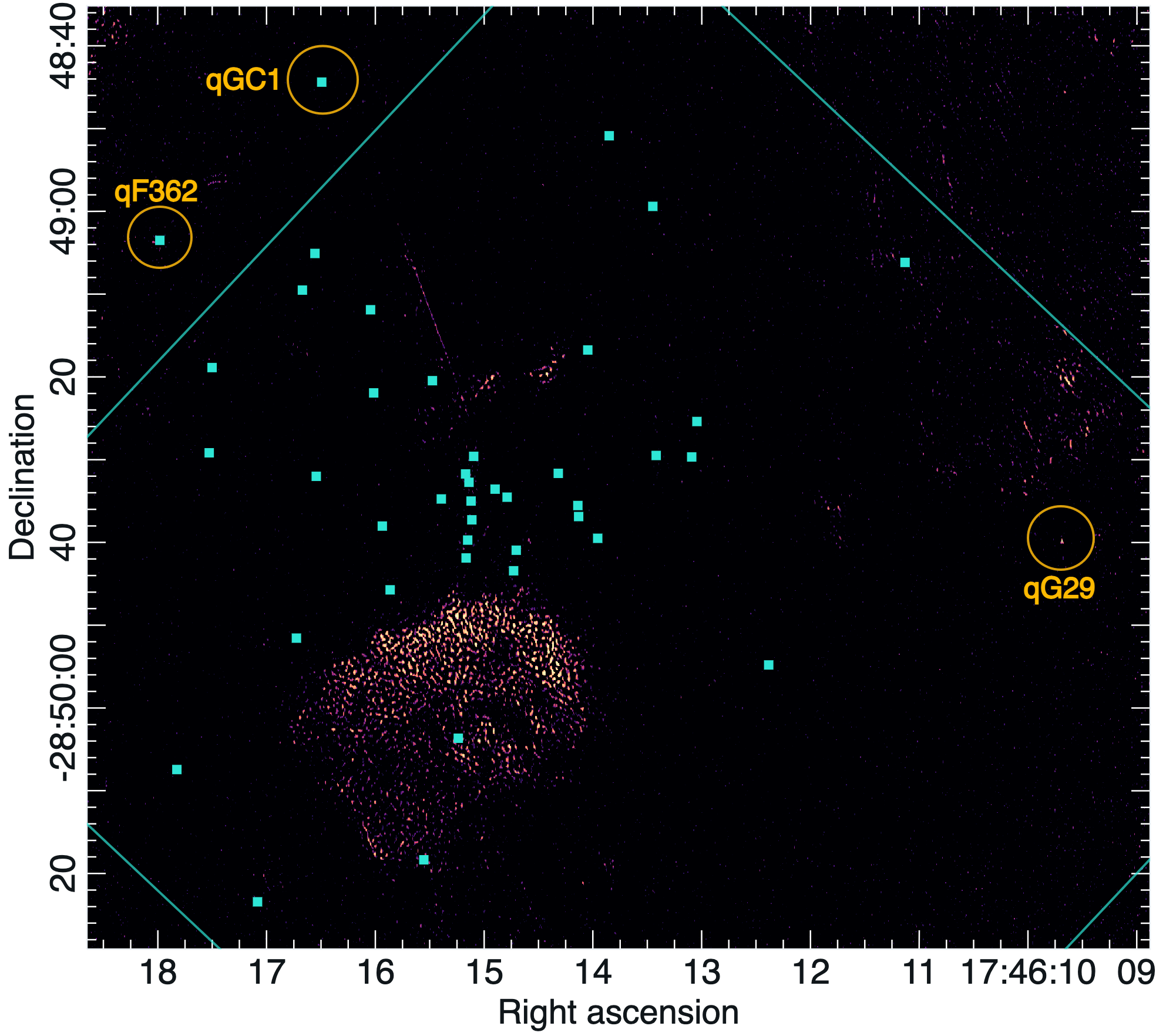}
     \caption{Positions of qF362, qCG1, and qG29 with respect to the Quintuplet cluster. The cyan squares represent the radio stars of Quintuplet. The cyan rhomboid shows the approximate area covered by \citet{Hosek2022}.}
        \label{Fig_pmsources}
\end{figure}

Unfortunately, qF362 and qCG1 fall outside of the area covered by \citet{Hosek2022}. However, we decided to keep them in the analysis because qF362 is in the \citet{Clark2018_Quintuplet} catalogue and because both of them are found within $\sim1\arcmin$ from the cluster centre. In addition, we looked for the proper motions of the source identified as qG29 in \citet{G-C2022}, which lies $>1\arcmin$ away from the cluster core. With their Bayesian analysis, \citet{Hosek2022} provide a zero cluster membership probability ($P_\mathrm{clust}=0.0$; see last row of Table \ref{Table_pm}), so we decided to exclude that detection from our analysis, as it most likely does not belong to Quintuplet\footnote{Furthermore, we could not find stellar counterparts for the \citet{G-C2022} sources qG5 and qG8 within the Pistol nebula. We believe that they are spurious contaminants and product of the over-resolved extended emission in the A-configuration.}. Figure \ref{Fig_pmsources} shows the locations of qF362, qCG1, and qG29 with respect to the Quintuplet cluster. In future work we will study the properties of the radio point sources scattered through the entire field of view of these observations.

\section{Discussion}\label{sect_discussion_conclusion}

\subsection{Spectral index variability}\label{subsect_alphavar}
\begin{figure}
  \centering
  \includegraphics[width=\hsize]{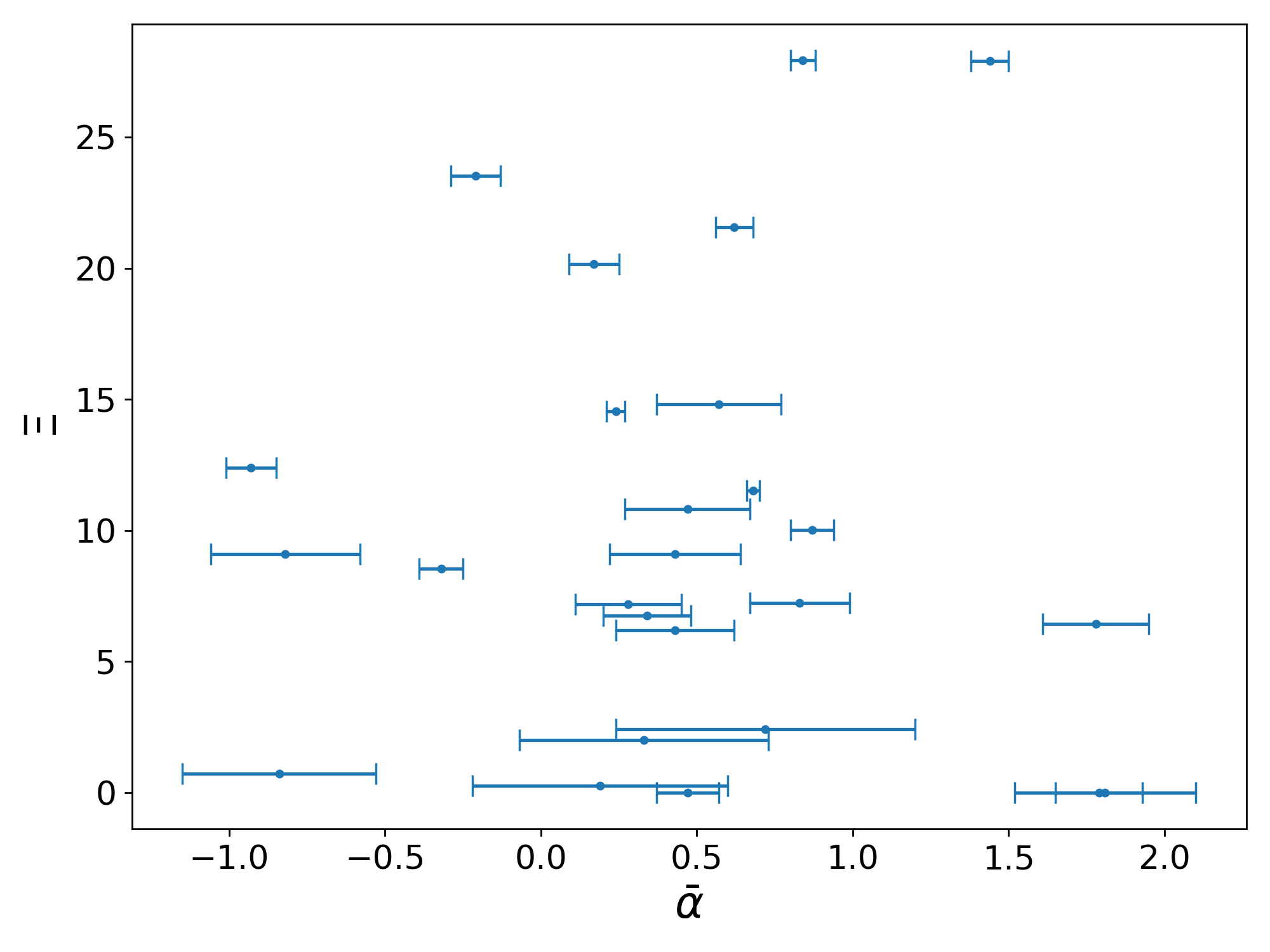}
     \caption{Spectral index variability versus weighted mean of the spectral index.}
        \label{Fig_Xi_vs_wmeanalpha}
\end{figure}

Variations in the spectral index of a radio star can indicate a change in the dominant emission mechanism (from thermal to non-thermal and vice versa). Table \ref{Table_var} shows the $\Xi$ values obtained for each radio-star, and Fig. \ref{Fig_Xi_vs_wmeanalpha} plots these values versus the weighted mean of the spectral index. 

\begin{figure}
  \centering
  \includegraphics[width=\hsize]{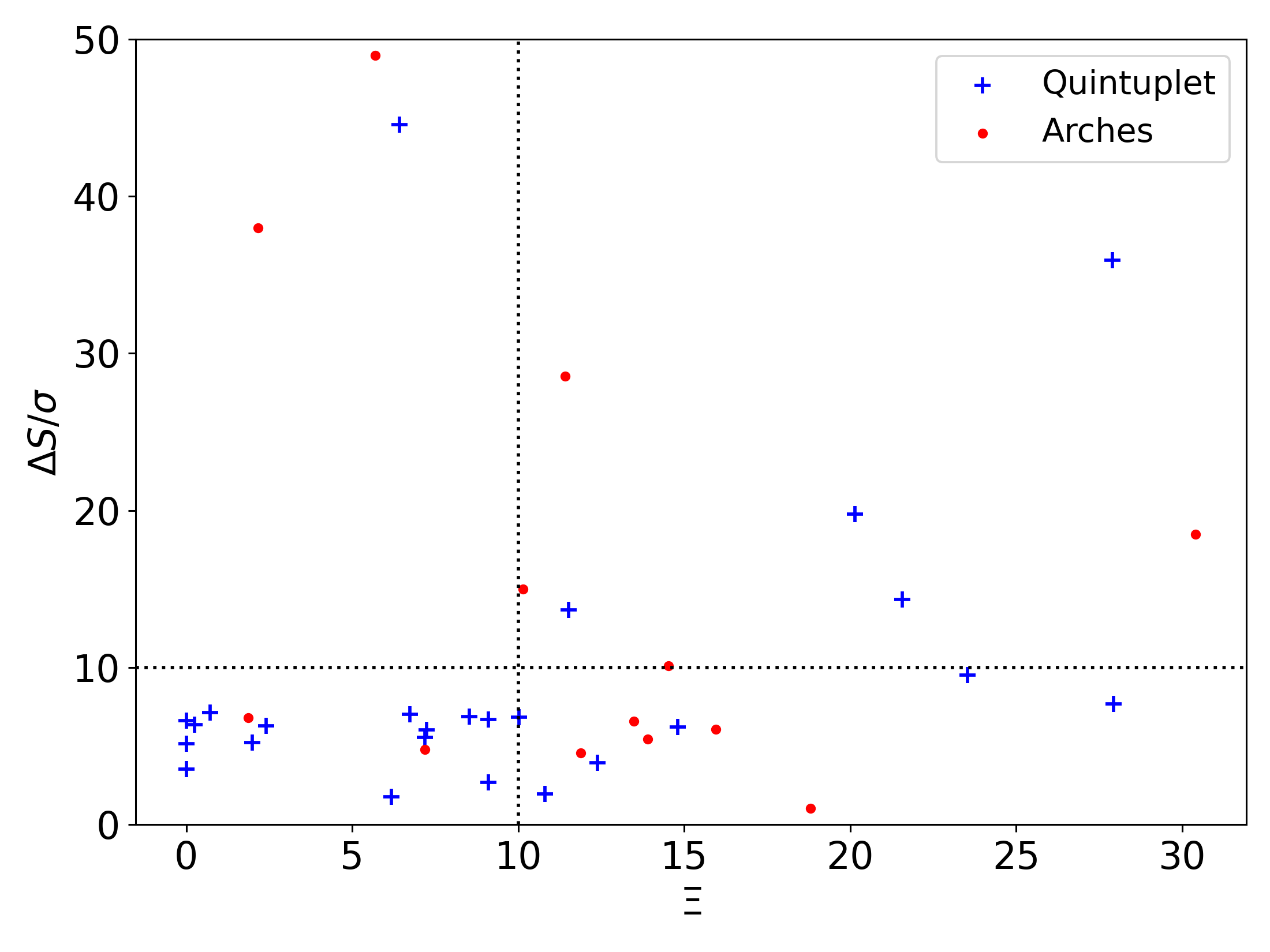}
     \caption{Flux density variability as a function of spectral index variability across all epochs for both the Arches and Quintuplet clusters. The black dotted lines indicate $\Delta S/\sigma = 10$ and $\Xi=10$ values, which are indicative of high flux and spectral index variability. Table \ref{Table_var} shows the uncertainties of the plotted values, but we refrained from showing error bars in this plot for the sake of clarity.}
        \label{Fig_fluxvar_vs_Xi}
\end{figure}

On the one hand, Fig. \ref{Fig_Xi_vs_wmeanalpha} shows that both thermal and non-thermal sources can showcase large spectral index variability. On the other hand, Fig. \ref{Fig_fluxvar_vs_Xi} shows that high flux density variability $(\Delta S/\sigma\gtrsim10)$ is frequently accompanied by a high variability of the spectral index $(\Xi\gtrsim10)$ across the time span of our observations. The only exception for this is qF134, the Pistol star $(\Delta S/\sigma=44.6\,, \Xi=6.4)$, whose variation in flux density is expected from a luminous blue variable (LBV).

\subsection{Mass-loss and clumping ratios}\label{subsect_Mdot}
We followed the same procedure as in \citet[][see their Eqs. 6 and 7]{Arches_paper} to derive the clumping-scaled mass-loss rates ($\dot{M}\sqrt{f_\mathrm{cl}}$, referred as $\dot{M}$ upper limits as well) from the thermal sources\footnote{The clumping factor is defined as $f_\mathrm{cl}=\langle \rho^2 \rangle/\langle \rho\rangle^2 \geq1$, with $f_\mathrm{cl}=1$ indicating a smooth, unclumped wind.}. In short, we used the \citet{Wright_Barlow1975} prescription, which assumes a fully ionised, isotropic stellar wind arising from a single star. 

For nearly all detections, we assumed a mean ionic charge of $Z=1\pm0.08$ (with the exception of the two LBVs, for which we used $Z=0.9\pm0.08$), a mean number of electrons per ion of $\gamma=1\pm0.08$ ($\gamma=0.8\pm0.08$ for the LBVs), and the electron temperature $T_e=10^4$ K. Also, we assumed a constant distance of 8 kpc to the GC.

We used terminal wind velocities ($v_\infty$) and mean molecular weight ($\mu$) values from previous literature. For WR stars of the WNh sub-type, we used the values from \citet{Liermann2010_WNh}, with the exception of qF241, qF240, and qF320, as they were given by Najarro (priv. comm.). All WR stars of the carbon sub-type (WC) use the same values as those provided by \citet{Najarro2017} for the pinwheel binary GCS4. For the rest of the OB giants not discussed in the aforementioned publications, we assumed $\mu=1.3$ and $v_\infty=1500\, \mathrm{km\,s^{-1}}$ ($v_\infty=500\, \mathrm{km\,s^{-1}}$), which are typical values for O-type (B-type) stars. We assumed a $10\%$ relative error for the terminal wind velocities in all radio stars. 

We searched for mass-loss variability across the time span of our observations by comparing upper limits of the mass-loss rates of each observing epoch, but we did not find any significant deviation from the mean value. Therefore, Table \ref{Table_Mdot} shows the adopted values for $v_\infty$, $\mu$, and the band-averaged, clumping-scaled mass-loss rates of the Quintuplet cluster radio stars.

\begin{table*}
\caption{Band-averaged $\dot{M}$ upper limits for the Quintuplet cluster radio stars.}
\label{Table_Mdot}
\centering
\begin{tabular}{l c c c c c c}
\hline \hline
ID & mu & $v_\infty$ & $\mathrm{avg}(\dot{M}_X)$ & $\mathrm{avg}(\dot{M}_C)$ & $\dot{M}_K$\tablefootmark{(a)} & reference\tablefootmark{(b)}   \\
\hline
 qF257 & 1.30 & 500  & --   & --   & $1.47 \pm 0.29$ & -- \\ 
 qF241 & 1.80 & 300  & $1.78 \pm 0.05$ & $1.74 \pm 0.05$ & $2.86 \pm 0.52$ & Najarro (priv. comm.) \\ 
 qF362 & 3.20 & 170  & $1.83 \pm 0.26$ & $1.36 \pm 0.05$ & -- & \citet{Najarro2009} \\ 
 qF240 & 1.80 & 400  & $1.29 \pm 0.04$ & $1.20 \pm 0.02$ & -- & Najarro (priv. comm.) \\ 
 qF320 & 2.50 & 700  & $3.22 \pm 0.11$ & $3.16 \pm 0.13$ & $4.58 \pm 0.83$ & Najarro (priv. comm.) \\ 
 qF211 & 4.70 & 1250 & $6.17 \pm 0.37$ & $5.94 \pm 0.50$ & -- & \citet{Najarro2017} \\ 
 qF256 & 2.60 & 900  & $2.51 \pm 0.24$ & $2.32 \pm 0.14$ & -- & \citet{Liermann2010_WNh} \\ 
 qF231 & 4.70 & 1250 & $5.60 \pm 0.31$ & $6.82 \pm 0.67$ & -- & \citet{Najarro2017} \\ 
qF270S & 3.50 & 300  & $0.96 \pm 0.07$ & $1.04 \pm 0.29$ & -- & -- \\ 
qF235S & 4.70 & 1250 & $5.13 \pm 0.13$ & $5.50 \pm 0.11$ & -- & \citet{Najarro2017} \\ 
qF235N & 4.70 & 1250 & $4.65 \pm 0.31$ & $5.14 \pm 0.48$ & -- & \citet{Najarro2017} \\ 
qF353E & 3.00 & 1600 & $3.59 \pm 0.28$ & $5.15 \pm 1.52$ & -- & \citet{Liermann2010_WNh} \\ 
 qF381 & 1.30 & 500  & $0.47 \pm 0.04$ & $0.55 \pm 0.02$ & -- & -- \\ 
 qF274 & 3.00 & 1250 & $2.09 \pm 0.10$ & $2.86 \pm 0.71$ & -- & \citet{Liermann2010_WNh} \\  
LHO076 & 4.70 & 1250 & $3.05 \pm 0.24$ & --   & -- & \citet{Najarro2017} \\ 
WR102ca & 4.70 &1250 & $2.94 \pm 0.35$ & --   & -- & \citet{Najarro2017} \\ 
 qF243 & 4.70 & 1250 & $2.57 \pm 0.38$ & --   & -- & \citet{Najarro2017} \\ 
 qF250 & 4.70 & 1250 & $2.42 \pm 0.50$ & --   & -- & \citet{Najarro2017} \\ 
LHO001 & 1.30 & 1500 & $0.73 \pm 0.10$ & --   & -- & -- \\ 
 qF258 & 4.70 & 1250 & $2.54 \pm 0.66$ & --   & -- & \citet{Najarro2017} \\ 
 qF251 & 4.70 & 1250 & $2.12 \pm 0.12$ & --   & -- & \citet{Najarro2017} \\ 
 qF134 & 2.80 & 105  & -- & -- & $3.63 \pm 0.96$ & \citet{Najarro2009} \\ 
\hline \hline
\end{tabular}
\tablefoot{All values and related uncertainties are in units of $10^{-5}\,M_\odot\,\mathrm{yr}^{-1}$. Errors represent the standard deviation. Terminal wind velocities in kilometres per second $(\mathrm{km\,s^{-1}})$ units. 
\tablefoottext{a}{Values derived from the K-band fluxes from \citet{Lang2005}.
\tablefoottext{b}{Sources with ' -- ' use typical $v_\infty$ and $\mu$ values for their spectral type (see text).}
}
}
\end{table*}

In order to remove non-thermal emitters from Table \ref{Table_Mdot}, we imposed that their weighted mean of the spectral index had to be larger than 0.3 ($\bar{\alpha}>0.3$, see values in Table \ref{Table_var}). As in \citet{Arches_paper}, we also used the K-band (central frequency of 22.5 GHz) fluxes from \citet{Lang2005} where available. This way, under the assumption that K-band flux is a better tracer of purely thermal radiation than our C- and X-band fluxes, we obtained an estimate of the $\dot{M}_K$ upper limit for qF257.

We note that since the two LBVs show spectral indices consistent with a wind affected by recombination ($\alpha\gtrsim1.3$) in most epochs, the assumption of a fully ionised wind does not hold, and the prescription by \citet{Wright_Barlow1975} is not valid. Therefore, following \citet{Panagia_Felli1975}, we derived the mass-loss prescription for a electron density profile of the form $n_e\propto r^{-3.5}$ (as expected from the winds of LBVs, e.g. \citealp{Cox1995}), and we obtained a flux density dependence of the form $S_\nu\propto\nu^{1.3}\,T_e^{0.55}\,\dot{M}^{2/3}\,v_\infty^{-2/3}\,R_*$. Thus, contrary to the fully ionised case, a detailed mass-loss formulation is now dependent on stellar radius, whose modelling (in the NIR) depends on the applied extinction law \citep[see, e.g.][]{Clark_IV}, leading to a source of uncertainty that is difficult to quantify with radio data alone. Therefore, in the cases in which the detected LBVs show $\alpha\gtrsim1.3$, we refrained from providing a value for $\dot{M}\sqrt{f_\mathrm{cl}}$ .

We also derived the clumping ratios $f_\mathrm{cl}^{\nu_1}/f_\mathrm{cl}^{\nu_2}$ $(\nu_2>\nu_1)$ as in \citet{Arches_paper}. Again, where available, we used the K-band fluxes from \citet{Lang2005} to compute $(f_\mathrm{cl}^C/f_\mathrm{cl}^K)$ and $(f_\mathrm{cl}^X/f_\mathrm{cl}^K)$, in which we took into account the fluxes from the deep C- and X-band images, respectively. Table \ref{table_clumpratios} shows the clumping ratios of the thermal sources ($\bar\alpha>0.3$).

\begin{table}
\caption{Clumping ratios}
\label{table_clumpratios}
\centering
\begin{tabular}{l c c c c}
\hline \hline
ID  & $(f_\mathrm{cl}^C/f_\mathrm{cl}^X)_{18}$ & $(f_\mathrm{cl}^C/f_\mathrm{cl}^X)_{22}$\tablefootmark{(a)} & $(f_\mathrm{cl}^C/f_\mathrm{cl}^K)$\tablefootmark{(b)} & $(f_\mathrm{cl}^X/f_\mathrm{cl}^K)$\tablefootmark{(b)} \\
\hline
  qF241 & $0.8 \pm 0.1$ & $0.9 \pm 0.1$ & $0.4 \pm 0.1$ & $0.4 \pm 0.1$  \\ 
  qF240 & $0.8 \pm 0.1$ & $0.8 \pm 0.1$ & -- & --  \\ 
  qF320 & $1.0 \pm 0.2$ & $0.9 \pm 0.2$ & $0.5 \pm 0.1$ & $0.5 \pm 0.1$  \\ 
  qF211 & $1.1 \pm 0.3$ & $0.8 \pm 0.2$ & -- & --  \\ 
  qF256 & $0.9 \pm 0.3$ & $0.8 \pm 0.2$ & -- & --  \\ 
  qF231 & $1.1 \pm 0.4$ & $1.4 \pm 0.4$ & -- & --  \\ 
 qF270S & $0.8 \pm 0.4$ & $0.8 \pm 0.3$ & -- & --  \\ 
 qF235S & $1.1 \pm 0.4$ & $1.0 \pm 0.3$ & -- & --  \\ 
 qF235N & $1.4 \pm 0.6$ & $1.3 \pm 0.6$ & -- & --  \\ 
 qF353E & -- & $1.1 \pm 0.5$ & -- & --  \\ 
  qF381 & -- & $1.3 \pm 0.5$ & -- & --  \\ 
  qF274 & $2.0 \pm 1.0$ & -- & -- & --  \\ 

\hline \hline
\end{tabular}
\tablefoot{
\tablefoottext{a}{Ratios computed using the combined C- and X-band data from 2022.}
\tablefoottext{b}{Ratios derived with K-band data from \citet{Lang2005} and the fluxes derived from our deep images.}
}
\end{table}

As can be seen in Table \ref{table_clumpratios}, most $(f_\mathrm{cl}^C/f_\mathrm{cl}^X)$ ratios approach unity, within the uncertainties. This indicates that both C- and X-band observations are similarly affected by clumping (at least radially outwards of the stellar photosphere). However, the clumping ratios derived with K-band fluxes show systematically lower values than unity, indicating that -- since K-band traces radiation closer to the stellar surface -- clumping becomes more relevant at higher frequencies (and therefore at inner wind regions, e.g. \citealt{Rubio-Diez2022}).

\subsection{Nature of the sources}
We discuss the nature of the detected point sources in this sub-section. We classify them into three broad categories based on their spectral indices and variability: CWB candidates, thermal sources, and ambiguous cases. Table \ref{Table_var} shows a summary of the different parameters used to classify the Quintuplet cluster radio stars. 

\subsubsection{Colliding-wind binary candidates}\label{subsect_CWB_candidates}
Following the approach of \citet{Arches_paper}, we may classify the radio stars into primary and secondary CWB candidates. In order for a radio-star to fall under the former category, it either needs to show a flat-to-negative spectral index consistently throughout the observations or show a negative spectral index with relatively low uncertainties ($\sigma_\alpha\lesssim0.2$) in at least one epoch. 

On the other hand, high flux density variability ($\Delta S/\sigma \gtrsim 10$), the presence of X-ray counterparts, and large changes in spectral index $(\Xi\gtrsim 10)$ may also indicate the presence of a companion \citep{DeBecker2013, Anastasopoulou2024}. Therefore, to fall under the secondary CWB category, a radio-star must either show high flux or spectral index variability, and/or present X-ray counterparts.

Following this approach, we confidently classified seven radio stars as primary CWB candidates: qF257, Muno7679, Dong66, qF344, qCG1, Hos332, and LHO090. We calssified four radio stars as secondary CWB candidates: qF320, qF211, qF256, and qF231. Thus, 11 out of 28 radio stars for which a spectral index could be obtained are classified as CWBs. Table \ref{Table_var} lists our CWB candidates, their maximum value of $\Delta S/\sigma$, a weighted mean value of the spectral indices, and their $\Xi$ parameter from Eq. (\ref{eq_alphavar}), which were used to discriminate their nature.

\begin{table*}
\caption{Variability parameters and source classification.}
\label{Table_var}
\centering
\begin{tabular}{lccccc}
\hline \hline
ID\tablefootmark{(a)}  &  $\max(\Delta S/\sigma)$\tablefootmark{(b)}  & $\bar{\alpha}\pm\sigma\bar{\alpha}$\tablefootmark{(c)} & $\Xi$\tablefootmark{(d)}  & stype\tablefootmark{(e)} & source classification \tablefootmark{(f)} \\
\hline
\textbf{qF257} & $143.3\pm4.9$ & $0.24 \pm 0.03$  &  $14.5\pm11.8$  & B1-2Ia+ & pCWB\\ 
qF241 & $13.7\pm3.9$ & $0.68 \pm 0.02$   &  $11.5\pm5.8$ & WN11h & Th.\\ 
qF362 & $35.9\pm2.9$ & $1.44 \pm 0.06$   &  $27.9\pm9.1$ & LBV & Th.\\ 
qF240 & $6.8\pm3.7$ & $0.87 \pm 0.07$  &  $10.0\pm8.4$ & WN10h & Th.\\ 
qF320 & $7.7\pm3.5$ & $0.84 \pm 0.04$  &   $27.9\pm9.3$ & WN9h & sCWB\\ 
\textbf{Muno7679} & $6.9\pm3.4$ & $-0.32 \pm 0.07$  & $8.5\pm5.0$ & -- & pCWB\\ 
Dong66 & $19.8\pm3.6$ & $0.12 \pm 0.08$  & $20.1\pm5.3$ & O7-B0 I & pCWB\\ 
\textbf{qF211} & $7.0\pm3.7$ & $0.34 \pm 0.14$  &  $6.7\pm3.5$ & WC9d(+OB) & sCWB\\ 
qF256 & $14.3\pm4.2$ & $0.62 \pm 0.06$  &   $21.6\pm12.6$ & WN8-9ha & sCWB\\ 
\textbf{qF231} & $6.0\pm3.2$ & $0.83 \pm 0.16$ &  $7.2\pm2.9$ & WC9d(+OB) & sCWB\\ 
qF270S & $6.7\pm3.1$ & $0.43 \pm 0.21$  &  $9.1\pm4.1$ & B1-2Ia+/WNLh & a\\ 
qF235S & $1.9\pm1.9$ & $0.47 \pm 0.20$  &  $10.8\pm4.8$ & WC8(d?+OB?) & a\\ 
qF307A & $5.6\pm3.5$ & $0.28 \pm 0.17$  &  $7.2\pm7.2$ & B1-2(Ia+) & a\\ 
\textbf{qF344} & $9.5\pm3.4$ & $-0.21 \pm 0.08$ & $23.5\pm7.8$ & O7-8Ia & pCWB\\ 
qF235N & $6.2\pm4.0$ & $0.57 \pm 0.20$  & $14.8\pm4.8$ & WC8(d?+OB?) & a\\ 
qF353E & $6.3\pm3.5$ & $0.72 \pm 0.48$  & $2.4\pm0.9$ & WN6 & a\\ 
qF381 & $5.2\pm2.3$ & $0.33 \pm 0.40$  & $2.0\pm1.4$ & B0-1Ia+/WNLh & a\\ 
qCG1 & $2.7\pm3.1$ & $-0.82 \pm 0.24$  & $9.1\pm3.7$ & -- & pCWB\\ 
qF278 & $4.4\pm3.4$ & $-0.23\pm0.58$   & -- & B0-1(Ia+) & a?\\ 
qF309 & $6.3\pm4.4$ & $0.19 \pm 0.41$  & $0.3\pm0.3$ & WC8-9(d?+OB?) & a\\ 
LHO100 & $7.9\pm3.3$ & $0.00\pm0.66$  & -- &  B2-3Ia+ & a\\ 
qF274 & $1.8\pm1.8$ & $0.43 \pm 0.19$   & $6.2\pm3.0$ & WN8-9ha & a? \\ 
Hos332 & $7.2\pm3.3$ & $-0.84 \pm 0.31$  & $0.7\pm0.6$ & -- & pCWB\\ 
LHO076 & $3.5\pm2.6$ & $0.47 \pm 0.10$  & -- & WC9d(+OB) & Th.\\ 
WR102ca & $6.6\pm4.5$ & $1.79 \pm 0.14$  & --& WC9d(+OB) & Th.\\ 
LHO090 & $3.9\pm3.1$ & $-0.93 \pm 0.08$   & $12.4\pm7.4$&  O9-B0Ia & pCWB\\ 
qF258 & $5.1\pm2.5$ & $1.81 \pm 0.29$   & --& WC9d(+OB) & Th.\\ 
qF134 & $44.6\pm3.3$ & $1.78 \pm 0.17$  & $6.4\pm2.6$ & LBV & Th.\\ 

\hline \hline
\end{tabular}
\tablefoot{\tablefoottext{a}{Boldface in ID indicates X-ray counterparts from \citet{Muno2009}.}
\tablefoottext{b}{Maximum value obtained from $\Delta S/\sigma$ in any band.}
\tablefoottext{c}{Spectral index's weighted mean.}
\tablefoottext{d}{Value obtained from Eq. \ref{eq_alphavar}. Uncertainties were derived using standard error propagation. }
\tablefoottext{e}{Spectral type from \citet{Clark2018_Quintuplet} or \citet{Dong2011}.}
\tablefoottext{f}{Source classification: pCWB = primary CWB candidate, sCWB = secondary CWB candidate, Th. = thermal source, a = ambiguous source. }
}
\end{table*}

\subsubsection{Thermal emitters}
Based on their positive $(\alpha\gtrsim0.6)$ spectral indices and relatively constant flux density across the timeline of our observations, we confidently classified seven sources as thermal emitters: qF241, qF362, qF240, LHO076, WR102ca, qF258, and qF134. 

We note that some of these thermal radio stars show $\alpha$ values considerably higher than the canonical $\alpha\approx0.6$ from \citet{Wright_Barlow1975,Panagia_Felli1975} for a fully ionised spherically symmetric wind.

In the case of the two LBVs (qF362 and qF134), highly thermal spectral index values ($S_\nu\propto \nu^{1.3}$) are expected from the steeper electron density profile characteristic of the winds of LBVs, which are affected by recombination (see Sect. \ref{subsect_Mdot}). However, the spectral indices derived for qF362 in some epochs (for example, the X-band observations from 4 October 2016) show even higher ($\alpha\gtrsim2$) values with relatively low uncertainties. This spectral behaviour could be a product of an even steeper $n_e$ profile presumably caused by inhomogeneities in the geometric structure of the wind (clumps). Nevertheless, we cannot exclude the possibility of a thermal contribution arising from the hot, dense optically thick colliding-wind region of a close (period of a few days) binary \citep[e.g.][]{Pittard2010,Montes2011}. The same explanations can be provided for the other thermal sources showing spectral indices higher than the canonical $\alpha\sim0.6$ value. Good examples non-LBV radio stars with $\alpha\gg0.6$ are qF240, WR102ca, and qF258. However, we do not consider high $\alpha$ values alone as a sufficient indicator of a close binary underlying their radio emission. Multi-epoch radial velocity studies of these radio stars would be an ideal complementary method to test whether close binaries may be the cause behind high $\alpha$ values.

In addition, one may expect that emission from optically thin winds could flatten the observed spectral index of a radio-star via purely thermal mechanisms. However, for mass-loss rates of $\gtrsim10^{-6}\, M_\odot\,\mathrm{yr^{-1}}$, this emission is traced at millimetre and sub-millimetre wavelengths (corresponding to frequencies of $\gtrsim100$ GHz; see Eq. 12 from \citealt{Wright_Barlow1975}). Furthermore, at these frequency regimes, the compact emission arising from the radio-star itself (which follows $S_\nu\propto\nu^{2}$) dominates over the contribution of the thin wind. For these reasons, we believe that deviations from $\alpha\sim0.6$ are predominantly caused by either a non-thermal contribution from a massive companion, inhomogeneities in the stellar wind, or a different electron density profile (or a combination of all).

\subsubsection{Ambiguous candidates}
Based on the individual spectral indices of each observation (particularly on the large uncertainties of some $\alpha$ values), on the weighted mean of the spectral index from Table \ref{Table_var}, and on the lack of data in some of our observations (only detected in some observations or the combined images), we are forced to conservatively classify ten sources as ambiguous. When studying their spectral index values from Table \ref{Table_alphas}, we note that most of these sources show compatibility with a non-thermal component in at least one epoch, always with considerable $\alpha$ uncertainties. The ten sources in question are qF270S, qF235S, qF307A, qF235N, qF235E, qF381, qF278, qF309, LHO100, and qF274.

\subsubsection{Multiplicity fraction}

Limiting ourselves to the 28 stars with measured spectral indices, the multiplicity fraction traced by plausible CWBs is $11/28=0.29\pm0.14$, assuming Poisson errors on the numbers. If we include the ambiguous cases, then the multiplicity fraction reaches $21/28=0.75\pm0.22$.

\subsubsection{Other sources and extended features}

Although this work is mainly focused on unresolved radio sources, the complexity of the radio environment surrounding the Quintuplet cluster unravels different extended structures that are worth mentioning, as they may hold a relation to the cluster and help clarify the nature of its stellar members.

\begin{figure}
  \centering
  \includegraphics[width=\hsize]{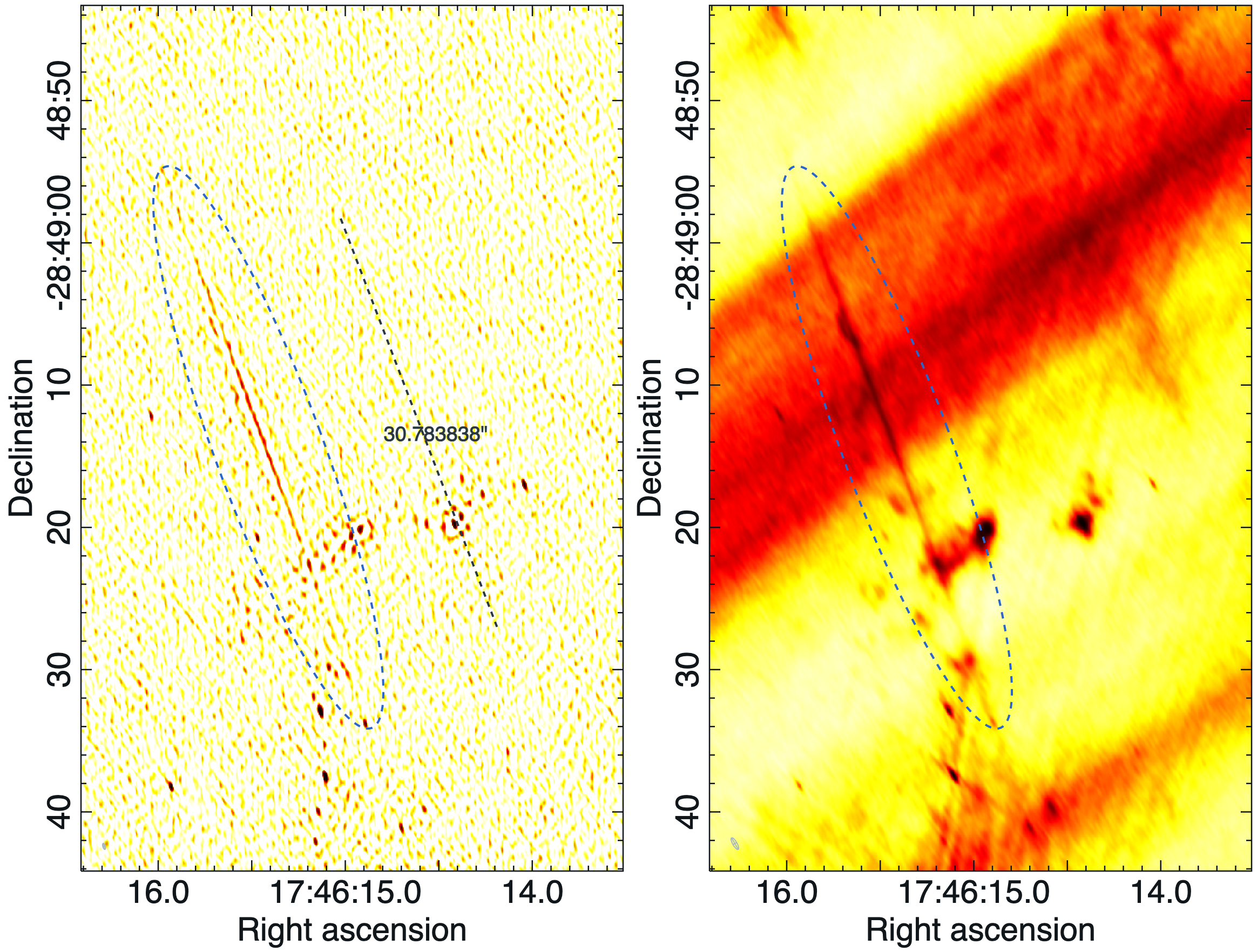}
     \caption{\textit{Left}: Position of the (non-thermal) thin filament in the Quintuplet deep C-band image (indicated by the blue ellipse). This image was created with the aforementioned \textit{u-v} cut. The black dotted line measures the apparent size of the filament. \textit{Right}: Same area as seen in the preliminary A, B, and C VLA multi-configuration C-band image (with no \textit{u-v} cut) in which both the thin filament and the larger non-thermal radio filament are clearly visible. The rms in the multi-configuration image is $\gtrsim50\,\mu$Jy/beam.}
        \label{Fig_filament}
\end{figure}

At all epochs, we find a thin filament that extends from approximately the centre of the Quintuplet cluster towards the north-east. This filament is also visible, but not discussed, in the images presented by \citep[][see their Fig.\,2]{G-C2022} . The filament is at least at least $30\arcsec$ ($\sim$1.2\,pc) long in projection but is unresolved transversally in both our C- and X-band observations; that is, the filament is thinner than $\sim0\farcs2$ ($\sim0.007$ pc). Given that the filament is more prominent in the C-band than in X-band, it appears to be dominated by non-thermal emission. The filament does not appear to be connected to any stellar source, and it does not appear to show the morphology of a jet emitted from, for instance, a microquasar or a massive protostar. It therefore appears to be a filament of the local magnetic field, illuminated by synchrotron radiation. Figure \ref{Fig_filament} shows the position of the filament with respect to the cluster in a preliminary multi-configuration image as well as our deep C-band image.

Non-thermal radio filaments are prominent all over the GC and are thought to trace $\sim$mG-strength magnetic fields oriented perpendicular to the Galactic Plane \citep[e.g.][]{Yusef-Zadeh1984,Morris1996,Heywood2022,Yusef-Zadeh2022}. However, so far, and to the best of our knowledge, no NTF has been reported near the Galactic Plane and oriented parallel to it, similar to the Quintuplet filament described here. Also, the Quintuplet filament is significantly shorter than the NTFs reported in the literature. We note that the Quintuplet NTF is oriented perpendicularly to the very prominent GC radio arc, that was recently studied in detail by \citet{Pare2019}. They did not report on this feature, probably because their angular resolution and sensitivity were not sufficient to pick it up. With knowledge of its existence, the Quintuplet NTF can (barely) be perceived in the MeerKAT image by \citet{Heywood2022}.

Because of its properties, we believe that the Quintuplet NTF is not related to the well-known large-scale NTFs that may trace the vertical magnetic field in the GC. While it is oriented parallel to the magnetic field orientation in the north-western part of the Sickle (Morris et al. in prep), the latter is dominated by thermal emission in filaments that are swept-up and somewhat deformed by the wind from the Quintuplet cluster. 

\begin{figure}
  \centering
  \includegraphics[width=\hsize]{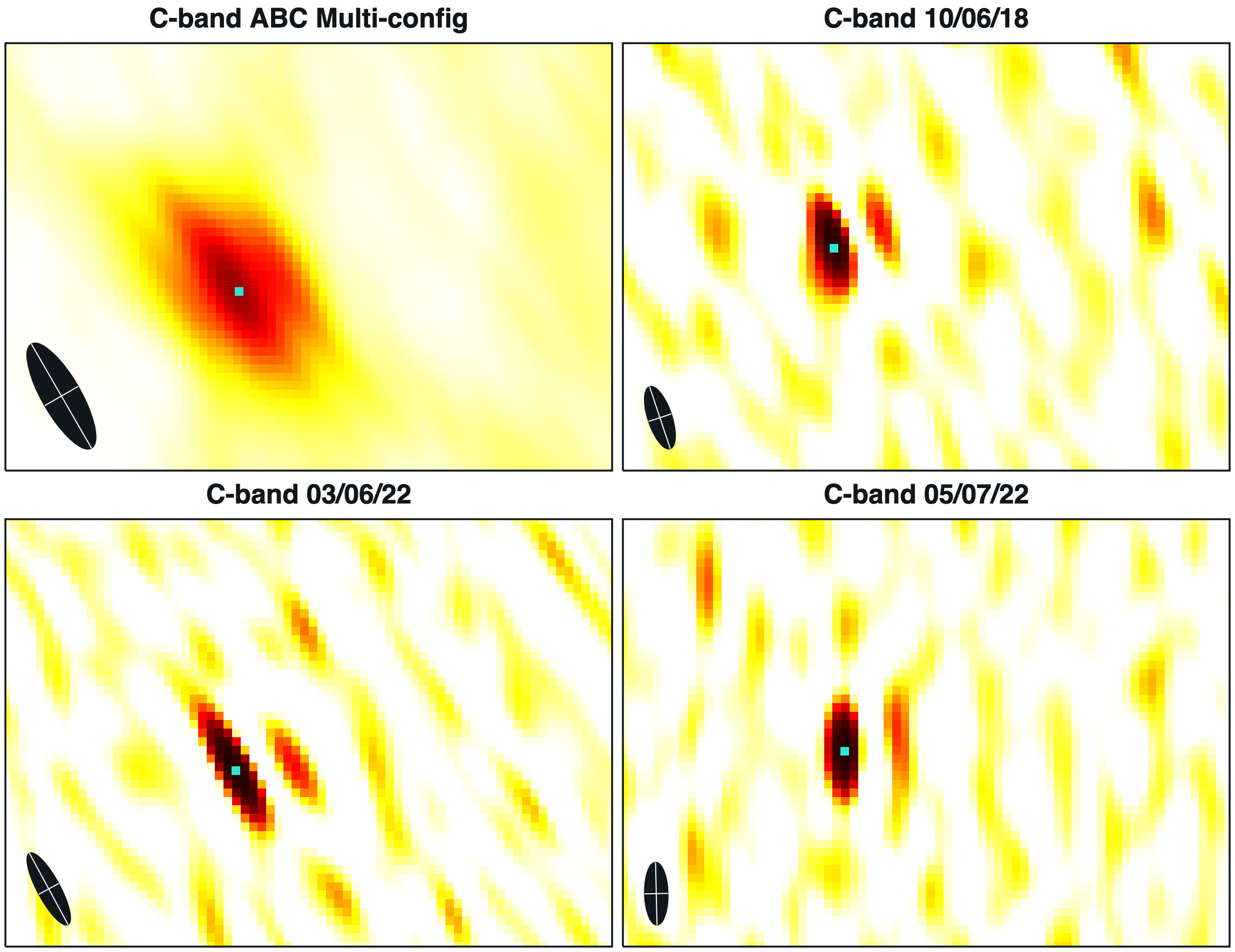}
     \caption{Close-up of the different C-band images of qF362, marked with the cyan dots.}
        \label{Fig_qF362closeup}
\end{figure}

Another feature that is worth mentioning is that we can clearly see hints of slightly extended emission next to the LBV qF362. Such blobby features next to the point source are more prominent in the C-band, which hints towards a non-thermal nature. Figure \ref{Fig_qF362closeup} shows a close-up of the LBV across our C-band observations.

Unfortunately, given the preliminary nature of the multi-configuration image and the over-resolution of the A-configuration, we cannot provide a quantitative analysis. Furthermore, if we estimate the angular size of the emitting envelope (assuming a fully ionised wind and purely thermal emission), as in \citet[][see their Eq. 26]{Panagia_Felli1975}, we obtain an angular size of $\theta_{ff}\sim0\farcs01$, which is a factor of $\sim30$ smaller than the C-band synthesised beam resolution in the A-configuration. Therefore, we deem it unlikely that the extended feature around qF362 is caused by the expanding envelope of the LBV, as contamination from the surrounding ionised medium (similar to the Sickle and Pistol nebulae) may be a more plausible explanation.

\subsection{Comparison with the Arches cluster}
When comparing the results presented here with our work with the Arches cluster \citep{Arches_paper}, we found some (mostly expected) similarities and differences.

We found a similar fraction of variable radio stars in both clusters, of $\sim 60\%$. However (perhaps with the exception of qF257), we could not find cases of extreme variability, as in the F18 and F26 radio stars of the Arches cluster, in which the emission drops below the detection limit on timescales of months. Given the random and relatively sparse time sampling combined with the unknown characteristics of the stars (e.g. binarity and corresponding orbital periods), we believe that it would be premature to draw any conclusions regarding the clusters themselves. Spectroscopic NIR follow-up observations of the highly variable radio stars could help constrain the orbital parameters of these presumable multiple systems.

\begin{figure}
  \centering
  \includegraphics[width=\hsize]{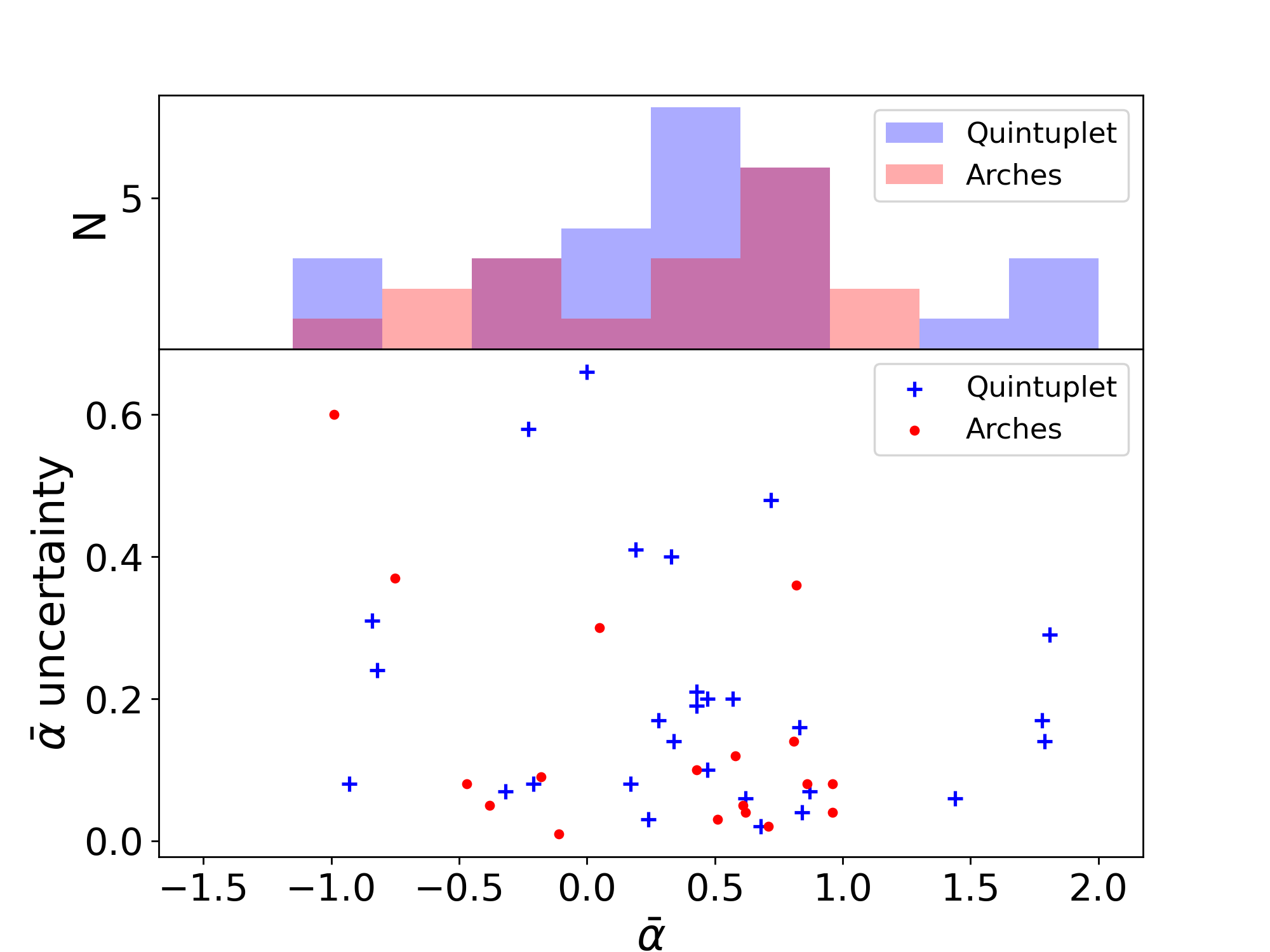}
     \caption{\textit{Top}: Histogram of the weighted mean of the spectral index for the radio stars of both clusters. Arches cluster data are in red; Quintuplet data are in blue. \textit{Bottom}: Uncertainty of the spectral index weighted mean versus spectral index weighted mean for both clusters. The values for Quintuplet are the same as those in Table \ref{Table_var}.}
        \label{Fig_walpha_hist_scatter}
\end{figure}

Figure \ref{Fig_walpha_hist_scatter} shows the histogram and scatter plot of the weighted mean of spectral index for both clusters. Although the histogram of Fig. \ref{Fig_walpha_hist_scatter} peaks at $\alpha\sim0.6$ for both clusters, indicating that the overall emission of the radio stars in the clusters is predominantly thermal, it is clear that the Quintuplet data spans a higher range in both absolute $\bar{\alpha}$ values and their uncertainties.

 On the one hand, we hypothesise that this behaviour is mostly due to instrumental effects, mainly because Quintuplet observations present an overall higher rms noise (specially in the C-band,  which is more sensitive to the bright non-thermal radio filament that lies across the Quintuplet field \citealp[see, e.g.][]{Pare2019}), resulting in higher uncertainties when deriving the spectral indices. On the other hand, the presence of two LBVs among the Quintuplet massive stellar population increases the $\alpha$ range towards highly thermal values $(\alpha>1)$. 
 
We also find that the radio stars of both clusters follow a trend of spectral index with stellar type. Namely, OB-type stars show that the weighted mean of their spectral indices is compatible with a non-thermal contribution to their emission. This is particularly evident for the Arches cluster, in which all O-type radio stars for which spectral indices were derived, clearly show $\alpha\lesssim0$. Table \ref{Table_OBtype_alpha} shows the $\bar\alpha$ values for the OB-type radio stars in the Arches and Quintuplet clusters.

\setlength{\tabcolsep}{15pt}
\begin{table}\label{Table_OBtype_alpha}
 \caption{Weighted mean of the spectral index for the OB-type radio stars of the Arches and Quintuplet clusters.}
\centering
\begin{tabular}{lcc}
\hline \hline
\multicolumn{3}{c}{Arches cluster} \\
\hline
ID\tablefootmark{(a)} & spectral type\tablefootmark{(a)} & $\bar\alpha$  \\
\hline
F19 & O4-5Ia & $-0.38 \pm 0.05$ \\
F18 & O4-5Ia+ & $-0.18 \pm 0.09$ \\
F26 & O4-5Ia & $-0.47 \pm 0.08$ \\
F55 & O5.5-6III & $-0.99\pm0.60$ \\
\hline

\multicolumn{3}{c}{Quintuplet cluster} \\

\hline
ID & spectral type\tablefootmark{(b)} & $\bar\alpha$  \\
\hline
qF257 & B1-2Ia+ & $0.24 \pm 0.03$ \\
Dong66 & O7-B0 & $0.17 \pm 0.08 $ \\
qF307A & B1-2(Ia+) & $0.28 \pm 0.17$ \\
qF344 & O7-8Ia & $-0.21 \pm 0.08$ \\
qF278 & B0-1(Ia+) & $-0.23 \pm 0.58$ \\
LHO100 & B2-3Ia+ & $0.00 \pm 0.66$ \\
LHO090 & O9-B0Ia & $-0.93 \pm 0.08$ \\
\hline\hline
\end{tabular}
\tablefoot{
\tablefoottext{a}{ID and spectral type from \citet{Clark_I}.}
\tablefoottext{b}{Spectral type from \citet{Clark2018_Quintuplet}.}
}
\end{table}

Another similarity in our studies of the Arches and Quintuplet clusters is that most of the radio stars that present high flux density variability (4/6 and 5/6 sources for Arches and Quintuplet, respectively, when setting $\Delta S/\sigma>10$ as threshold) also show large changes of spectral index during the time span of our observations $(\Xi\gtrsim10)$.

Considering that the evolutionary state of the radio stars of both clusters plays a crucial role in the characterisation of the wind parameters and in the derived clumping-scaled mass-loss rates, we can only compare the $\dot{M}\sqrt{f_\mathrm{cl}}$ factors of the stars that share spectral type between the clusters, that is, WNh7-9 type stars. We find that the $\dot{M}\sqrt{f_\mathrm{cl}}$ factors for the Arches WNh stars span a larger range $(1-5\times10^{-5}\, M_\odot\,\mathrm{yr^{-1}})$ than those of the Quintuplet WNh stars $(2-3\times10^{-5}\, M_\odot\,\mathrm{yr^{-1}})$. This larger range may be explained by the fact that Arches hosts a larger WNh sample than Quintuplet with a total of 15 WNh7-9 stars, 12 more than Quintuplet. Also, since Arches is roughly 1 Myr younger than Quintuplet, it is more likely that the most massive stars, which present stronger mass-loss, are still alive.

If we restrict our analysis to the radio realm only, both young massive clusters show similar multiplicity fractions (Arches: $39\%$, Quintuplet: $\sim27\%$). However, the multiplicity fraction of the Arches radio stars increases to $\sim60\%$ when taking into account NIR radial velocity studies, and at the time of writing, no analogous spectroscopic studies have been carried out on the Quintuplet cluster. Nevertheless, if we consider ambiguous detections as potential CWB candidates, the Quintuplet multiplicity fraction increases to $\sim75\%$. Multi-wavelength diagnostics of binarity (e.g. radio-continuum studies, radial velocity NIR studies, deeper X-ray observations) are sensitive to different orbital separations of multiple systems and would provide a more complete census of binaries in the GC young massive clusters.

\section{Summary and conclusions}\label{sect_conclusions}

We have presented the most complete radio-stellar catalogue of the Quintuplet cluster to date, with a total of 41 radio point sources associated with stellar NIR counterparts, increasing the number of radio stars by $\sim50\%$ when compared to the most recent study.

We cross-matched the Quintuplet radio point sources with recent proper motion catalogues to ensure their cluster membership. We find that most of the detected radio stars show a high cluster membership probability, according to the Bayesian study of \citet{Hosek2022}.

We find that around $60\%$ of the radio stars of the Quintuplet cluster show variability in their radio emission on timescales of years, a value that is shared with our recent analogous study of the Arches cluster.

We determined the spectral indices for 28 of 41 sources. Based on their values, and flux and/or spectral index variability, we classified 11 sources as CWB candidates, seven as thermal emitters, and ten sources as ambiguous, mostly due to their high spectral index uncertainties. Upcoming instruments, such as the SKAO-MID radio interferometer, will certainly shed light on the ambiguous nature of some of the radio stars with their improved observational capabilities.

Thus, using the criterion of non-thermal emission alone, we find a $\approx27\%$ multiplicity fraction of the Quintuplet radio stars. However, this value is most certainly a lower limit, considering the ambiguous nature of many of our radio stars, and the lack of multi-frequency (e.g. NIR radial velocity) studies on the Quintuplet cluster at the time of writing. When taking into account ambiguous sources, the multiplicity fraction of the Quintuplet radio stars increases to $\sim75\%$.

We derived the upper limits (scaled by the clumping factor) of the mass-loss rates of the Quintuplet thermal radio stars. We find that they range from $\sim1-6\times10^{-5}\, M_\odot\, \mathrm{yr^{-1}}$, consistent with values from WNh, WC, and OB supergiant stars. 

We also computed the clumping ratios of the thermally emitting radio stars. We find that, as was the case for the Arches cluster, the C- and X-band observations are similarly affected by clumping, but clumping may become stronger at higher frequencies (such as at the K-band). More multi-frequency (e.g. millimetre and sub-millimetre, centimetre, NIR), multi-facility -- for example ALMA, (ng)VLA, SKAO-MID, ELT -- observations of the GC Arches and Quintuplet clusters would help sample clumping at different atmospheric heights, and could help understand the role it plays during the massive stellar evolution of post main-sequence stars.

When comparing the Arches and Quintuplet radio stars, we find that the average spectral index of both clusters falls in canonically thermal values. However, the Quintuplet radio stars show a larger scatter in both mean values and uncertainties, presumably because of instrumental factors. Furthermore, with the current sensitivity of the  VLA, we find that the OB-type radio stars of both clusters are more likely to show compatibility with a non-thermal component in their emission.

\begin{acknowledgements} 

We dearly thank the NRAO staff for their guidance during the observations and calibration process.

MCG, RS, AA, JM, MPT, and ATGC acknowledge financial support from the Severo Ochoa grant CEX2021-001131-S funded by MCIN/AEI/ 10.13039/501100011033. MGC and RS acknowledge support from grant EUR2022-134031 funded by MCIN/AEI/10.13039/501100011033 and by the European Union NextGenerationEU/PRTR and by grant PID2022- 136640NB-C21 funded by MCIN/AEI 10.13039/501100011033 and by the European Union.
Authors AA, JM and MPT acknowledge financial support from the Spanish grant PID2023-147883NB-C21, funded by MCIU/AEI/ 10.13039/501100011033, as well as support through ERDF/EU.

ATGC acknowledges the Astrophysics and High Energy Physics programme supported by MCIN with funding from European Union NextGenerationEU (PRTR-C17.I1) and by Generalitat Valenciana.
AA acknowledges support by the PID2023-147883NB-C21 grant.

F.N., acknowledges support by grant PID2022-137779OB-C41 funded by
MCIN/AEI/10.13039/501100011033 by "ERDF A way of making Europe".

Authors MCG, RS and JM acknowledge the Spanish Prototype of an SRC (SPSRC) service and support funded by the Ministerio de Ciencia, Innovación y Universidades (MICIU), by the Junta de Andalucía, by the European Regional Development Funds (ERDF) and by the European Union NextGenerationEU/PRTR. The SPSRC acknowledges financial support from the Agencia Estatal de Investigación (AEI) through the "Center of Excellence Severo Ochoa" award to the Instituto de Astrofísica de Andalucía (IAA-CSIC) (SEV-2017-0709) and from the grant CEX2021-001131-S funded by MICIU/AEI/ 10.13039/501100011033.

JM acknowledges financial support from the grant  PID2021-123930OB-C21 funded by MICIU/AEI/ 10.13039/501100011033 and by ERDF/EU.

\end{acknowledgements}

\bibliographystyle{aa} 
\bibliography{references_Q} 

\begin{thebibliography}{46}
\expandafter\ifx\csname natexlab\endcsname\relax\def\natexlab#1{#1}\fi

\bibitem[{{Anastasopoulou} {et~al.}(2024){Anastasopoulou}, {Guarcello}, {Flaccomio}, {Sciortino}, {Benatti}, {De Becker}, {Wright}, {Drake}, {Albacete-Colombo}, {Andersen}, {Argiroffi}, {Bayo}, {Castellanos}, {Gennaro}, {Grebel}, {Miceli}, {Najarro}, {Negueruela}, {Prisinzano}, {Ritchie}, {Robberto}, {Sabbi}, \& {Zeidler}}]{Anastasopoulou2024}
{Anastasopoulou}, K., {Guarcello}, M.~G., {Flaccomio}, E., {et~al.} 2024, \aap, 690, A25

\bibitem[{{Cano-Gonz{\'a}lez} {et~al.}(2024){Cano-Gonz{\'a}lez}, {Sch{\"o}del}, {Alberdi}, {Mold{\'o}n}, {P{\'e}rez-Torres}, {Najarro}, \& {Gallego-Calvente}}]{Arches_paper}
{Cano-Gonz{\'a}lez}, M., {Sch{\"o}del}, R., {Alberdi}, A., {et~al.} 2024, \aap, 692, A23

\bibitem[{{CASA Team} {et~al.}(2022){CASA Team}, {Bean}, {Bhatnagar}, {Castro}, {Donovan Meyer}, {Emonts}, {Garcia}, {Garwood}, {Golap}, {Gonzalez Villalba}, {Harris}, {Hayashi}, {Hoskins}, {Hsieh}, {Jagannathan}, {Kawasaki}, {Keimpema}, {Kettenis}, {Lopez}, {Marvil}, {Masters}, {McNichols}, {Mehringer}, {Miel}, {Moellenbrock}, {Montesino}, {Nakazato}, {Ott}, {Petry}, {Pokorny}, {Raba}, {Rau}, {Schiebel}, {Schweighart}, {Sekhar}, {Shimada}, {Small}, {Steeb}, {Sugimoto}, {Suoranta}, {Tsutsumi}, {van Bemmel}, {Verkouter}, {Wells}, {Xiong}, {Szomoru}, {Griffith}, {Glendenning}, \& {Kern}}]{CASA}
{CASA Team}, {Bean}, B., {Bhatnagar}, S., {et~al.} 2022, \pasp, 134, 114501

\bibitem[{Castor {et~al.}(1975)Castor, Abbott, \& Klein}]{Castor_Abbot_Klein1975}
Castor, J.~I., Abbott, D.~C., \& Klein, R.~I. 1975, Astrophysical Journal, vol. 195, Jan. 1, 1975, pt. 1, p. 157-174., 195, 157

\bibitem[{{Clark} {et~al.}(2018{\natexlab{a}}){Clark}, {Lohr}, {Najarro}, {Dong}, \& {Martins}}]{Clark_I}
{Clark}, J.~S., {Lohr}, M.~E., {Najarro}, F., {Dong}, H., \& {Martins}, F. 2018{\natexlab{a}}, \aap, 617, A65

\bibitem[{{Clark} {et~al.}(2023){Clark}, {Lohr}, {Najarro}, {Patrick}, \& {Ritchie}}]{Clark_IV}
{Clark}, J.~S., {Lohr}, M.~E., {Najarro}, F., {Patrick}, L.~R., \& {Ritchie}, B.~W. 2023, \mnras, 521, 4473

\bibitem[{{Clark} {et~al.}(2018{\natexlab{b}}){Clark}, {Lohr}, {Patrick}, {Najarro}, {Dong}, \& {Figer}}]{Clark2018_Quintuplet}
{Clark}, J.~S., {Lohr}, M.~E., {Patrick}, L.~R., {et~al.} 2018{\natexlab{b}}, \aap, 618, A2

\bibitem[{{Condon}(1997)}]{Condon1997}
{Condon}, J.~J. 1997, \pasp, 109, 166

\bibitem[{{Cox} {et~al.}(1995){Cox}, {Mezger}, {Sievers}, {Najarro}, {Bronfman}, {Kreysa}, \& {Haslam}}]{Cox1995}
{Cox}, P., {Mezger}, P.~G., {Sievers}, A., {et~al.} 1995, \aap, 297, 168

\bibitem[{{De Becker}(2007)}]{DeBecker2007}
{De Becker}, M. 2007, \aapr, 14, 171

\bibitem[{{De Becker} \& {Raucq}(2013)}]{DeBecker2013}
{De Becker}, M. \& {Raucq}, F. 2013, \aap, 558, A28

\bibitem[{{Do} {et~al.}(2013){Do}, {Lu}, {Ghez}, {Morris}, {Yelda}, {Martinez}, {Wright}, \& {Matthews}}]{Do2013}
{Do}, T., {Lu}, J.~R., {Ghez}, A.~M., {et~al.} 2013, \apj, 764, 154

\bibitem[{{Dong} {et~al.}(2011){Dong}, {Wang}, {Cotera}, {Stolovy}, {Morris}, {Mauerhan}, {Mills}, {Schneider}, {Calzetti}, \& {Lang}}]{Dong2011}
{Dong}, H., {Wang}, Q.~D., {Cotera}, A., {et~al.} 2011, \mnras, 417, 114

\bibitem[{{Dougherty} {et~al.}(2005){Dougherty}, {Beasley}, {Claussen}, {Zauderer}, \& {Bolingbroke}}]{Dougherty2005}
{Dougherty}, S.~M., {Beasley}, A.~J., {Claussen}, M.~J., {Zauderer}, B.~A., \& {Bolingbroke}, N.~J. 2005, \apj, 623, 447

\bibitem[{Figer {et~al.}(1999)Figer, Kim, Morris, Serabyn, Rich, \& McLean}]{Figer1999}
Figer, D.~F., Kim, S.~S., Morris, M., {et~al.} 1999, The Astrophysical Journal, 525, 750

\bibitem[{{Gallego-Calvente} {et~al.}(2021){Gallego-Calvente}, {Sch{\"o}del}, {Alberdi}, {Herrero-Illana}, {Najarro}, {Yusef-Zadeh}, {Dong}, {Sanchez-Bermudez}, {Shahzamanian}, {Nogueras-Lara}, \& {Gallego-Cano}}]{G-C2021}
{Gallego-Calvente}, A.~T., {Sch{\"o}del}, R., {Alberdi}, A., {et~al.} 2021, \aap, 647, A110

\bibitem[{{Gallego-Calvente} {et~al.}(2022){Gallego-Calvente}, {Sch{\"o}del}, {Alberdi}, {Najarro}, {Yusef-Zadeh}, {Shahzamanian}, \& {Nogueras-Lara}}]{G-C2022}
{Gallego-Calvente}, A.~T., {Sch{\"o}del}, R., {Alberdi}, A., {et~al.} 2022, \aap, 664, A49

\bibitem[{{GRAVITY Collaboration} {et~al.}(2019){GRAVITY Collaboration}, {Abuter}, {Amorim}, {Baub{\"o}ck}, {Berger}, {Bonnet}, {Brandner}, {Cl{\'e}net}, {Coud{\'e} Du Foresto}, {de Zeeuw}, {Dexter}, {Duvert}, {Eckart}, {Eisenhauer}, {F{\"o}rster Schreiber}, {Garcia}, {Gao}, {Gendron}, {Genzel}, {Gerhard}, {Gillessen}, {Habibi}, {Haubois}, {Henning}, {Hippler}, {Horrobin}, {Jim{\'e}nez-Rosales}, {Jocou}, {Kervella}, {Lacour}, {Lapeyr{\`e}re}, {Le Bouquin}, {L{\'e}na}, {Ott}, {Paumard}, {Perraut}, {Perrin}, {Pfuhl}, {Rabien}, {Rodriguez Coira}, {Rousset}, {Scheithauer}, {Sternberg}, {Straub}, {Straubmeier}, {Sturm}, {Tacconi}, {Vincent}, {von Fellenberg}, {Waisberg}, {Widmann}, {Wieprecht}, {Wiezorrek}, {Woillez}, \& {Yazici}}]{GRAVITY2019}
{GRAVITY Collaboration}, {Abuter}, R., {Amorim}, A., {et~al.} 2019, \aap, 625, L10

\bibitem[{{Heywood} {et~al.}(2022){Heywood}, {Rammala}, {Camilo}, {Cotton}, {Yusef-Zadeh}, {Abbott}, {Adam}, {Adams}, {Aldera}, {Asad}, {Bauermeister}, {Bennett}, {Bester}, {Bode}, {Botha}, {Botha}, {Brederode}, {Buchner}, {Burger}, {Cheetham}, {de Villiers}, {Dikgale-Mahlakoana}, {du Toit}, {Esterhuyse}, {Fanaroff}, {February}, {Fourie}, {Frank}, {Gamatham}, {Geyer}, {Goedhart}, {Gouws}, {Gumede}, {Hlakola}, {Hokwana}, {Hoosen}, {Horrell}, {Hugo}, {Isaacson}, {J{\'o}zsa}, {Jonas}, {Joubert}, {Julie}, {Kapp}, {Kenyon}, {Kotz{\'e}}, {Kriek}, {Kriel}, {Krishnan}, {Lehmensiek}, {Liebenberg}, {Lord}, {Lunsky}, {Madisa}, {Magnus}, {Mahgoub}, {Makhaba}, {Makhathini}, {Malan}, {Manley}, {Marais}, {Martens}, {Mauch}, {Merry}, {Millenaar}, {Mnyandu}, {Mokone}, {Monama}, {Mphego}, {New}, {Ngcebetsha}, {Ngoasheng}, {Ockards}, {Oozeer}, {Otto}, {Passmoor}, {Patel}, {Peens-Hough}, {Perkins}, {Ramaila}, {Ramanujam}, {Ramudzuli}, {Ratcliffe}, {Robyntjies}, {Salie}, {Sambu}, {Schollar}, {Schwardt}, {Schwartz}, {Serylak},
  {Siebrits}, {Sirothia}, {Slabber}, {Smirnov}, {Sofeya}, {Taljaard}, {Tasse}, {Tiplady}, {Toruvanda}, {Twum}, {van Balla}, {van der Byl}, {van der Merwe}, {Van Tonder}, {Van Wyk}, {Venter}, {Venter}, {Wallace}, {Welz}, {Williams}, \& {Xaia}}]{Heywood2022}
{Heywood}, I., {Rammala}, I., {Camilo}, F., {et~al.} 2022, \apj, 925, 165

\bibitem[{{H{\"o}gbom}(1974)}]{Hogbom74}
{H{\"o}gbom}, J.~A. 1974, \aaps, 15, 417

\bibitem[{{Hosek} {et~al.}(2022){Hosek}, {Do}, {Lu}, {Morris}, {Ghez}, {Martinez}, \& {Anderson}}]{Hosek2022}
{Hosek}, M.~W., {Do}, T., {Lu}, J.~R., {et~al.} 2022, \apj, 939, 68

\bibitem[{Lang {et~al.}(2005)Lang, Johnson, Goss, \& Rodriguez}]{Lang2005}
Lang, C., Johnson, K., Goss, W., \& Rodriguez, L. 2005, The Astronomical Journal, 130

\bibitem[{Liermann {et~al.}(2010)Liermann, Hamann, Oskinova, Todt, \& Butler}]{Liermann2010_WNh}
Liermann, A., Hamann, W.-R., Oskinova, L., Todt, H., \& Butler, K. 2010, Astronomy \& Astrophysics, 524, A82

\bibitem[{{Liermann} {et~al.}(2012){Liermann}, {Hamann}, \& {Oskinova}}]{Liermann2012}
{Liermann}, A., {Hamann}, W.~R., \& {Oskinova}, L.~M. 2012, \aap, 540, A14

\bibitem[{Ludovici {et~al.}(2016)Ludovici, Lang, Morris, Mutel, Mills, IV, \& Ott}]{Ludovici2016}
Ludovici, D.~A., Lang, C.~C., Morris, M.~R., {et~al.} 2016, The Astrophysical Journal, 826, 218

\bibitem[{{Marchant} \& {Bodensteiner}(2024)}]{Marchant_Bodensteiner2024}
{Marchant}, P. \& {Bodensteiner}, J. 2024, \araa, 62, 21

\bibitem[{Montes {et~al.}(2011)Montes, Gonz{\'a}lez, Cant{\'o}, P{\'e}rez-Torres, \& Alberdi}]{Montes2011}
Montes, G., Gonz{\'a}lez, R., Cant{\'o}, J., P{\'e}rez-Torres, M., \& Alberdi, A. 2011, Astronomy \& Astrophysics, 531, A52

\bibitem[{Morris \& Serabyn(1996)}]{Morris1996}
Morris, M. \& Serabyn, E. 1996, Annual Review of Astronomy and Astrophysics, 34, 645

\bibitem[{{Muno} {et~al.}(2009){Muno}, {Bauer}, {Baganoff}, {Bandyopadhyay}, {Bower}, {Brandt}, {Broos}, {Cotera}, {Eikenberry}, {Garmire}, {Hyman}, {Kassim}, {Lang}, {Lazio}, {Law}, {Mauerhan}, {Morris}, {Nagata}, {Nishiyama}, {Park}, {Ram{\`\i}rez}, {Stolovy}, {Wijnands}, {Wang}, {Wang}, \& {Yusef-Zadeh}}]{Muno2009}
{Muno}, M.~P., {Bauer}, F.~E., {Baganoff}, F.~K., {et~al.} 2009, \apjs, 181, 110

\bibitem[{Najarro {et~al.}(2009)Najarro, Figer, Hillier, Geballe, \& Kudritzki}]{Najarro2009}
Najarro, F., Figer, D.~F., Hillier, D.~J., Geballe, T., \& Kudritzki, R.~P. 2009, The Astrophysical Journal, 691, 1816

\bibitem[{Najarro {et~al.}(2017)Najarro, Geballe, Figer, \& de~la Fuente}]{Najarro2017}
Najarro, F., Geballe, T., Figer, D., \& de~la Fuente, D. 2017, The Astrophysical Journal, 845, 127

\bibitem[{Panagia \& Felli(1975)}]{Panagia_Felli1975}
Panagia, N. \& Felli, M. 1975, Astronomy and Astrophysics, vol. 39, no. 1, Feb. 1975, p. 1-5., 39, 1

\bibitem[{{Par{\'e}} {et~al.}(2019){Par{\'e}}, {Lang}, {Morris}, {Moore}, \& {Mao}}]{Pare2019}
{Par{\'e}}, D.~M., {Lang}, C.~C., {Morris}, M.~R., {Moore}, H., \& {Mao}, S.~A. 2019, \apj, 884, 170

\bibitem[{Paumard {et~al.}(2006)Paumard, Genzel, Martins, Nayakshin, Beloborodov, Levin, Trippe, Eisenhauer, Ott, Gillessen, {et~al.}}]{Paumard2006}
Paumard, T., Genzel, R., Martins, F., {et~al.} 2006, The Astrophysical Journal, 643, 1011

\bibitem[{{Perley} \& {Butler}(2017)}]{Perley_Butler2017}
{Perley}, R.~A. \& {Butler}, B.~J. 2017, \apjs, 230, 7

\bibitem[{{Pittard}(2010)}]{Pittard2010}
{Pittard}, J.~M. 2010, \mnras, 403, 1633

\bibitem[{{Rau} \& {Cornwell}(2011)}]{Rau_&_Cornwell2011}
{Rau}, U. \& {Cornwell}, T.~J. 2011, \aap, 532, A71

\bibitem[{Rubio-D{\'\i}ez {et~al.}(2022)Rubio-D{\'\i}ez, Sundqvist, Najarro, Traficante, Puls, Calzoletti, \& Figer}]{Rubio-Diez2022}
Rubio-D{\'\i}ez, M., Sundqvist, J., Najarro, F., {et~al.} 2022, Astronomy \& Astrophysics, 658, A61

\bibitem[{{Sana} {et~al.}(2012){Sana}, {de Mink}, {de Koter}, {Langer}, {Evans}, {Gieles}, {Gosset}, {Izzard}, {Le Bouquin}, \& {Schneider}}]{Sana2012}
{Sana}, H., {de Mink}, S.~E., {de Koter}, A., {et~al.} 2012, Science, 337, 444

\bibitem[{{Sana} \& {Evans}(2011)}]{Sana2011}
{Sana}, H. \& {Evans}, C.~J. 2011, in IAU Symposium, Vol. 272, Active OB Stars: Structure, Evolution, Mass Loss, and Critical Limits, ed. C.~{Neiner}, G.~{Wade}, G.~{Meynet}, \& G.~{Peters}, 474--485

\bibitem[{{Sanchez-Bermudez} {et~al.}(2019){Sanchez-Bermudez}, {Alberdi}, {Sch{\"o}del}, {Brandner}, {Galv{\'a}n-Madrid}, {Guirado}, {Herrero-Illana}, {Hummel}, {Marcaide}, \& {P{\'e}rez-Torres}}]{Sanchez-Bermudez2019}
{Sanchez-Bermudez}, J., {Alberdi}, A., {Sch{\"o}del}, R., {et~al.} 2019, \aap, 624, A55

\bibitem[{{Thompson} {et~al.}(2017){Thompson}, {Moran}, \& {Swenson}}]{Thompson2017}
{Thompson}, A.~R., {Moran}, J.~M., \& {Swenson}, George~W., J. 2017, {Interferometry and Synthesis in Radio Astronomy, 3rd Edition}

\bibitem[{{Wright} \& {Barlow}(1975)}]{Wright_Barlow1975}
{Wright}, A.~E. \& {Barlow}, M.~J. 1975, \mnras, 170, 41

\bibitem[{Yusef-Zadeh {et~al.}(2022)Yusef-Zadeh, Arendt, Wardle, Heywood, \& Cotton}]{Yusef-Zadeh2022}
Yusef-Zadeh, F., Arendt, R.~G., Wardle, M., Heywood, I., \& Cotton, W. 2022, Monthly Notices of the Royal Astronomical Society, 517, 294

\bibitem[{Yusef-Zadeh {et~al.}(1984)Yusef-Zadeh, Morris, \& Chance}]{Yusef-Zadeh1984}
Yusef-Zadeh, F., Morris, M., \& Chance, D. 1984, Nature, 310, 557

\bibitem[{{Zhao} {et~al.}(2020){Zhao}, {Morris}, \& {Goss}}]{Zhao2020}
{Zhao}, J.-H., {Morris}, M.~R., \& {Goss}, W.~M. 2020, \apj, 905, 173

\end{thebibliography}

\begin{appendix}
\section{Additional material}

\begin{table*}
\caption{Observations and image properties.}
\label{table_observations}      
\centering
\begin{tabular}{l c c c c}
\hline \hline
Date  &  Band & ($\theta_{\mathrm{maj}}^\mathrm{FWHM}\times\theta_{\mathrm{min}}^{\mathrm{FWHM}},\ \mathrm{PA}$)  & Off-source rms ($\mu$Jy/beam)& Time on target (minutes)\\    
\hline                        
4 Oct 2016  & X & $0\farcs46\times0\farcs15,\ 30\degr$ & $6.4$ &  52\\      
27 Oct 2016 & X & $0\farcs38\times0\farcs14,\ 18\degr$  & $5.4$ & 47\\
combined 2016  & X & $0\farcs38\times0\farcs14,\ 21\degr$ & $4.5$ & 99\\
24 Mar 2018  & X & $0\farcs36\times0\farcs14,\ -7\degr$ & $5.3$ &  47\\      
10 Jun 2018  & C & $0\farcs51\times0\farcs19,\ 18\degr$  & $7.6$ & 76\\
20 May 2022    & X & $0\farcs41\times0\farcs13,\ 25\degr$ & $4.6$ &  86\\     
2 Jul 2022   & X & $0\farcs35\times0\farcs14,\ 14\degr$   & $5.7$ & 62\\
4 Jul 2022   & X & $0\farcs41\times0\farcs14,\ -24\degr$   & $5.9$ & 62\\
combined 2022   & X & $0\farcs28\times0\farcs14,\ 10\degr$     & $3.1$ & 210\\
3 Jun 2022  & C & $0\farcs63\times0\farcs18,\ 28\degr$    & $7.8$ & 86\\
5 Jul 2022  & C & $0\farcs49\times0\farcs18,\ 1\degr$    & $7.8$ & 62\\
combined 2022   & C & $0\farcs44\times0\farcs18,\ 14\degr$     & $5.9$ & 148\\

\hline
Deep X-band   & X & $0\farcs29\times0\farcs14,\ 10\degr$     & $2.3$ & 356\\
Deep C-band   & C & $0\farcs44\times0\farcs18,\ 15\degr$     & $5.6$ & 224\\
\hline \hline

\end{tabular}
\tablefoot{All data were acquired with A-configuration.}
\end{table*}

\def\arraystretch{1.25}
\begin{sidewaystable*}
\footnotesize
\begin{center}
\caption{Quintuplet radio point sources}
\label{fluxtable}
\resizebox{1.0\textwidth}{!}{
\begin{tabular}{lcccccccccccccccccc}
\hline 
\vspace{2pt}
 ID  & R.A. (\degr) & eR.A. (\arcsec) & $\delta$ (\degr) &  e$\delta$ (\arcsec) & $S_X^{\mathrm{deep}}$ &  $S_C^{\mathrm{deep}}$ & $S_X^{04/10/16}$ & $S_X^{27/10/16}$ &  $S_X^{2016}$ & $S_X^{24/03/18}$ & $S_X^{20/05/22}$ & $S_X^{02/07/22}$ & $S_X^{04/07/22}$ & $S_X^{2022}$ & $S_C^{2018}$ & $S_C^{03/06/22}$ & $S_C^{05/07/22}$ & $S_C^{2022}$ \\ 
\hline
 qF257 & 266.563078 & 0.007 & -28.825759 & 0.01 & 0.503 $\pm$ 0.021 & 0.338 $\pm$ 0.024 & 1.007 $\pm$ 0.033 & 1.088 $\pm$ 0.039 & 1.072 $\pm$ 0.036 & 0.525 $\pm$ 0.021 & 0.290 $\pm$ 0.016 & 0.299 $\pm$ 0.015 & 0.260 $\pm$ 0.020 & 0.283 $\pm$ 0.014 & 0.454 $\pm$ 0.029 & 0.248 $\pm$ 0.025 & 0.209 $\pm$ 0.019 & 0.226 $\pm$ 0.024 \\
  qF241 & 266.562971 & 0.007 & -28.827023 & 0.01 & 0.485 $\pm$ 0.019 & 0.355 $\pm$ 0.022 & 0.512 $\pm$ 0.019 & 0.487 $\pm$ 0.020 & 0.511 $\pm$ 0.018 & 0.538 $\pm$ 0.021 & 0.477 $\pm$ 0.018 & 0.496 $\pm$ 0.020 & 0.469 $\pm$ 0.020 & 0.489 $\pm$ 0.018 & 0.354 $\pm$ 0.025 & 0.341 $\pm$ 0.027 & 0.373 $\pm$ 0.028 & 0.343 $\pm$ 0.026 \\
 qF362 & 266.574915 & 0.01 & -28.817639 & 0.01 & 0.366 $\pm$ 0.017 & 0.185 $\pm$ 0.017 & 0.449 $\pm$ 0.024 & 0.395 $\pm$ 0.021 & 0.410 $\pm$ 0.020 & 0.535 $\pm$ 0.021 & 0.346 $\pm$ 0.019 & 0.303 $\pm$ 0.016 & 0.315 $\pm$ 0.019 & 0.322 $\pm$ 0.016 & 0.179 $\pm$ 0.020 & 0.201 $\pm$ 0.021 & 0.189 $\pm$ 0.019 & 0.182 $\pm$ 0.019 \\
  qF240 & 266.566402 & 0.007 & -28.827228 & 0.01 & 0.214 $\pm$ 0.009 & 0.143 $\pm$ 0.013 & 0.236 $\pm$ 0.013 & 0.210 $\pm$ 0.013 & 0.225 $\pm$ 0.010 & 0.234 $\pm$ 0.011 & 0.213 $\pm$ 0.011 & 0.218 $\pm$ 0.013 & 0.208 $\pm$ 0.012 & 0.219 $\pm$ 0.010 & 0.147 $\pm$ 0.016 & 0.145 $\pm$ 0.018 & 0.151 $\pm$ 0.017 & 0.141 $\pm$ 0.015 \\
  qF320 & 266.558530 & 0.008 & -28.821322 & 0.01 & 0.210 $\pm$ 0.009 & 0.159 $\pm$ 0.015 & 0.234 $\pm$ 0.013 & 0.211 $\pm$ 0.012 & 0.226 $\pm$ 0.010 & 0.234 $\pm$ 0.012 & 0.226 $\pm$ 0.012 & 0.217 $\pm$ 0.013 & 0.203 $\pm$ 0.011 & 0.225 $\pm$ 0.010 & 0.173 $\pm$ 0.019 & 0.154 $\pm$ 0.015 & 0.153 $\pm$ 0.017 & 0.157 $\pm$ 0.018 \\
Muno7679 & 266.574271 & 0.01 & -28.835401 & 0.01 & 0.172 $\pm$ 0.007 & 0.209 $\pm$ 0.014 & 0.190 $\pm$ 0.013 & 0.164 $\pm$ 0.012 & 0.183 $\pm$ 0.009 & 0.192 $\pm$ 0.012 & 0.168 $\pm$ 0.011 & 0.185 $\pm$ 0.012 & 0.163 $\pm$ 0.011 & 0.175 $\pm$ 0.010 & 0.222 $\pm$ 0.017 & 0.210 $\pm$ 0.018 & 0.201 $\pm$ 0.016 & 0.207 $\pm$ 0.017 \\
 Dong66 & 266.564813 & 0.01 & -28.838427 & 0.02 & 0.109 $\pm$ 0.008 & 0.094 $\pm$ 0.022 & 0.089 $\pm$ 0.012 & 0.092 $\pm$ 0.011 & 0.096 $\pm$ 0.010 & 0.176 $\pm$ 0.014 & 0.087 $\pm$ 0.010 & 0.112 $\pm$ 0.011 & 0.093 $\pm$ 0.012 & 0.098 $\pm$ 0.009 & 0.121 $\pm$ 0.031 & -- & 0.104 $\pm$ 0.026 & 0.056 $\pm$ 0.016 \\
  qF211 & 266.566100 & 0.008 & -28.829370 & 0.01 & 0.104 $\pm$ 0.006 & 0.083 $\pm$ 0.011 & 0.123 $\pm$ 0.013 & 0.102 $\pm$ 0.011 & 0.108 $\pm$ 0.009 & 0.103 $\pm$ 0.011 & 0.109 $\pm$ 0.010 & 0.116 $\pm$ 0.010 & 0.093 $\pm$ 0.010 & 0.110 $\pm$ 0.007 & 0.082 $\pm$ 0.014 & 0.064 $\pm$ 0.015 & 0.082 $\pm$ 0.015 & 0.071 $\pm$ 0.012 \\
  qF256 & 266.568932 & 0.008 & -28.825556 & 0.01 & 0.099 $\pm$ 0.005 & 0.067 $\pm$ 0.010 & 0.140 $\pm$ 0.013 & 0.098 $\pm$ 0.010 & 0.115 $\pm$ 0.008 & 0.091 $\pm$ 0.008 & 0.108 $\pm$ 0.007 & 0.096 $\pm$ 0.008 & 0.108 $\pm$ 0.010 & 0.109 $\pm$ 0.007 & 0.064 $\pm$ 0.013 & 0.076 $\pm$ 0.017 & 0.076 $\pm$ 0.015 & 0.069 $\pm$ 0.012 \\
  qF231 & 266.561268 & 0.007 & -28.828036 & 0.01 & 0.090 $\pm$ 0.006 & 0.088 $\pm$ 0.016 & 0.097 $\pm$ 0.012 & 0.089 $\pm$ 0.011 & 0.096 $\pm$ 0.009 & 0.099 $\pm$ 0.009 & 0.089 $\pm$ 0.009 & 0.109 $\pm$ 0.010 & 0.085 $\pm$ 0.009 & 0.094 $\pm$ 0.007 & 0.077 $\pm$ 0.018 & 0.106 $\pm$ 0.021 & 0.091 $\pm$ 0.019 & 0.087 $\pm$ 0.016 \\
 qF270S & 266.562896 & 0.007 & -28.824890 & 0.01 & 0.088 $\pm$ 0.007 & 0.066 $\pm$ 0.018 & 0.088 $\pm$ 0.011 & 0.091 $\pm$ 0.012 & 0.098 $\pm$ 0.010 & 0.095 $\pm$ 0.009 & 0.070 $\pm$ 0.009 & 0.095 $\pm$ 0.010 & 0.090 $\pm$ 0.009 & 0.085 $\pm$ 0.007 & 0.063 $\pm$ 0.020 & 0.113 $\pm$ 0.027 & 0.047 $\pm$ 0.015 & 0.053 $\pm$ 0.012 \\
 qF235S & 266.563188 & 0.007 & -28.828299 & 0.01 & 0.079 $\pm$ 0.005 & 0.065 $\pm$ 0.011 & 0.086 $\pm$ 0.010 & 0.085 $\pm$ 0.012 & 0.089 $\pm$ 0.009 & 0.087 $\pm$ 0.012 & 0.083 $\pm$ 0.010 & 0.079 $\pm$ 0.009 & 0.082 $\pm$ 0.010 & 0.084 $\pm$ 0.006 & 0.070 $\pm$ 0.013 & 0.069 $\pm$ 0.018 & 0.066 $\pm$ 0.015 & 0.061 $\pm$ 0.013 \\
 qF307A & 266.564484 & 0.008 & -28.822352 & 0.01 & 0.085 $\pm$ 0.005 & 0.074 $\pm$ 0.010 & 0.087 $\pm$ 0.012 & 0.081 $\pm$ 0.010 & 0.088 $\pm$ 0.008 & 0.093 $\pm$ 0.010 & 0.074 $\pm$ 0.008 & 0.095 $\pm$ 0.011 & 0.084 $\pm$ 0.011 & 0.086 $\pm$ 0.007 & 0.084 $\pm$ 0.015 & 0.065 $\pm$ 0.013 & 0.064 $\pm$ 0.014 & 0.064 $\pm$ 0.010 \\
  qF344 & 266.569461 & 0.009 & -28.819314 & 0.01 & 0.071 $\pm$ 0.005 & 0.094 $\pm$ 0.012 & 0.070 $\pm$ 0.011 & 0.051 $\pm$ 0.010 & 0.054 $\pm$ 0.008 & 0.065 $\pm$ 0.008 & 0.088 $\pm$ 0.009 & 0.077 $\pm$ 0.010 & 0.074 $\pm$ 0.010 & 0.088 $\pm$ 0.007 & 0.064 $\pm$ 0.015 & 0.133 $\pm$ 0.017 & 0.119 $\pm$ 0.017 & 0.124 $\pm$ 0.015 \\
 qF235N & 266.563126 & 0.008 & -28.827700 & 0.01 & 0.067 $\pm$ 0.007 & 0.069 $\pm$ 0.014 & 0.089 $\pm$ 0.012 & 0.064 $\pm$ 0.011 & 0.075 $\pm$ 0.009 & 0.078 $\pm$ 0.010 & 0.071 $\pm$ 0.010 & 0.072 $\pm$ 0.010 & 0.073 $\pm$ 0.009 & 0.071 $\pm$ 0.008 & 0.073 $\pm$ 0.020 & 0.054 $\pm$ 0.018 & 0.061 $\pm$ 0.017 & 0.061 $\pm$ 0.017 \\
 qF353E & 266.546378 & 0.01 & -28.818377 & 0.02 & 0.063 $\pm$ 0.005 & 0.053 $\pm$ 0.014 & 0.067 $\pm$ 0.011 & 0.086 $\pm$ 0.011 & 0.071 $\pm$ 0.009 & 0.069 $\pm$ 0.009 & 0.066 $\pm$ 0.010 & 0.063 $\pm$ 0.009 & 0.061 $\pm$ 0.009 & 0.066 $\pm$ 0.007 & -- & -- & 0.082 $\pm$ 0.028 & 0.051 $\pm$ 0.015 \\
  qF381 & 266.556037 & 0.01 & -28.816500 & 0.02 & 0.063 $\pm$ 0.005 & 0.053 $\pm$ 0.010 & 0.072 $\pm$ 0.011 & 0.072 $\pm$ 0.011 & 0.080 $\pm$ 0.009 & 0.067 $\pm$ 0.008 & 0.064 $\pm$ 0.009 & 0.062 $\pm$ 0.010 & 0.052 $\pm$ 0.009 & 0.062 $\pm$ 0.007 & -- & 0.064 $\pm$ 0.015 & 0.057 $\pm$ 0.014 & 0.056 $\pm$ 0.012 \\
    qCG1 & 266.568725 & 0.01 & -28.812327 & 0.02 & 0.053 $\pm$ 0.005 & 0.079 $\pm$ 0.012 & 0.053 $\pm$ 0.011 & 0.053 $\pm$ 0.011 & 0.057 $\pm$ 0.009 & 0.052 $\pm$ 0.009 & 0.059 $\pm$ 0.009 & 0.052 $\pm$ 0.010 & 0.053 $\pm$ 0.010 & 0.058 $\pm$ 0.007 & 0.093 $\pm$ 0.015 & 0.084 $\pm$ 0.020 & 0.111 $\pm$ 0.022 & 0.082 $\pm$ 0.013 \\
  qF278 & 266.562999 & 0.008 & -28.826394 & 0.02 & 0.047 $\pm$ 0.006 & 0.043 $\pm$ 0.012 & 0.045 $\pm$ 0.015 & 0.043 $\pm$ 0.010 & 0.044 $\pm$ 0.008 & 0.053 $\pm$ 0.011 & 0.035 $\pm$ 0.009 & 0.044 $\pm$ 0.009 & 0.045 $\pm$ 0.010 & 0.041 $\pm$ 0.007 & -- & -- & -- & 0.046 $\pm$ 0.011 \\
  qF309 & 266.572922 & 0.01 & -28.821914 & 0.01 & 0.044 $\pm$ 0.004 & 0.038 $\pm$ 0.009 & 0.054 $\pm$ 0.012 & 0.045 $\pm$ 0.008 & 0.050 $\pm$ 0.006 & 0.050 $\pm$ 0.008 & 0.050 $\pm$ 0.007 & 0.034 $\pm$ 0.007 & 0.044 $\pm$ 0.008 & 0.046 $\pm$ 0.005 & 0.043 $\pm$ 0.012 & 0.058 $\pm$ 0.014 & 0.047 $\pm$ 0.016 & 0.043 $\pm$ 0.012 \\
    LHO100 & 266.563211 & 0.01 & -28.825477 & 0.03 & 0.054 $\pm$ 0.013 & 0.064 $\pm$ 0.016 & 0.049 $\pm$ 0.011 & 0.051 $\pm$ 0.012 & 0.051 $\pm$ 0.010 & 0.033 $\pm$ 0.008 & 0.069 $\pm$ 0.013 & 0.063 $\pm$ 0.015 & 0.038 $\pm$ 0.012 & 0.057 $\pm$ 0.011 & -- & -- & -- & 0.057 $\pm$ 0.016 \\
  qF274 & 266.573026 & 0.009 & -28.824771 & 0.01 & 0.044 $\pm$ 0.004 & 0.049 $\pm$ 0.011 & 0.044 $\pm$ 0.012 & 0.049 $\pm$ 0.009 & 0.051 $\pm$ 0.007 & 0.045 $\pm$ 0.009 & 0.043 $\pm$ 0.006 & 0.043 $\pm$ 0.008 & 0.048 $\pm$ 0.009 & 0.048 $\pm$ 0.006 & 0.052 $\pm$ 0.014 & -- & -- & -- \\
    Hos332 & 266.566847 & 0.009 & -28.819970 & 0.01 & 0.042 $\pm$ 0.004 & 0.065 $\pm$ 0.011 & 0.032 $\pm$ 0.008 & 0.035 $\pm$ 0.009 & 0.039 $\pm$ 0.008 & 0.040 $\pm$ 0.007 & 0.056 $\pm$ 0.008 & 0.054 $\pm$ 0.010 & 0.038 $\pm$ 0.009 & 0.050 $\pm$ 0.006 & 0.078 $\pm$ 0.017 & 0.095 $\pm$ 0.016 & 0.057 $\pm$ 0.012 & 0.073 $\pm$ 0.014 \\
    LHO076 & 266.558906 & 0.008 & -28.826541 & 0.01 & 0.038 $\pm$ 0.004 & 0.037 $\pm$ 0.009 & 0.048 $\pm$ 0.009 & 0.044 $\pm$ 0.010 & 0.050 $\pm$ 0.008 & 0.036 $\pm$ 0.007 & 0.039 $\pm$ 0.008 & 0.040 $\pm$ 0.007 & 0.041 $\pm$ 0.011 & 0.040 $\pm$ 0.006 & -- & -- & -- & -- \\
WR102ca & 266.554348 & 0.010 & -28.823716 & 0.02 & 0.037 $\pm$ 0.004 & -- & 0.054 $\pm$ 0.013 & 0.040 $\pm$ 0.009 & 0.045 $\pm$ 0.007 & 0.034 $\pm$ 0.006 & 0.038 $\pm$ 0.009 & 0.041 $\pm$ 0.009 & 0.031 $\pm$ 0.009 & 0.039 $\pm$ 0.007 & -- & -- & -- & -- \\
    LHO090 & 266.562079 & 0.007 & -28.825994 & 0.02 & 0.032 $\pm$ 0.004 & 0.079 $\pm$ 0.012 & -- & -- & -- & 0.037 $\pm$ 0.009 & 0.040 $\pm$ 0.008 & 0.053 $\pm$ 0.011 & 0.036 $\pm$ 0.008 & 0.049 $\pm$ 0.006 & 0.098 $\pm$ 0.022 & 0.089 $\pm$ 0.017 & 0.068 $\pm$ 0.014 & 0.080 $\pm$ 0.014 \\
  qF243 & 266.558878 & 0.008 & -28.826915 & 0.02 & 0.028 $\pm$ 0.004 & -- & 0.039 $\pm$ 0.010 & 0.026 $\pm$ 0.006 & 0.031 $\pm$ 0.006 & 0.023 $\pm$ 0.007 & 0.040 $\pm$ 0.008 & 0.039 $\pm$ 0.009 & -- & 0.038 $\pm$ 0.006 & -- & -- & -- & -- \\
    Hos18 & 266.554551 & 0.01 & -28.824915 & 0.02 & 0.025 $\pm$ 0.004 & -- & -- & -- & 0.037 $\pm$ 0.008 & -- & -- & 0.033 $\pm$ 0.009 & -- & 0.029 $\pm$ 0.005 & -- & -- & -- & -- \\
  qF250 & 266.564142 & 0.009 & -28.826321 & 0.02 & 0.021 $\pm$ 0.004 & -- & -- & -- & -- & -- & -- & -- & -- & 0.031 $\pm$ 0.006 & -- & -- & -- & -- \\
    LHO001 & 266.569694 & 0.01 & -28.830988 & 0.02 & 0.020 $\pm$ 0.004 & -- & -- & -- & 0.027 $\pm$ 0.007 & -- & 0.029 $\pm$ 0.007 & 0.034 $\pm$ 0.010 & -- & 0.020 $\pm$ 0.005 & -- & -- & -- & -- \\
  qF258 & 266.559657 & 0.008 & -28.825457 & 0.02 & 0.025 $\pm$ 0.003 & -- & -- & -- & -- & 0.045 $\pm$ 0.010 & -- & -- & -- & 0.022 $\pm$ 0.005 & -- & -- & -- & -- \\
  qF358 & 266.568978 & 0.01 & -28.818084 & 0.02 & 0.018 $\pm$ 0.004 & -- & -- & -- & -- & -- & -- & -- & -- & 0.023 $\pm$ 0.005 & -- & -- & -- & -- \\
  qF251 & 266.561624 & 0.008 & -28.826255 & 0.02 & 0.017 $\pm$ 0.003 & -- & -- & -- & 0.028 $\pm$ 0.007 & -- & -- & -- & -- & 0.024 $\pm$ 0.007 & -- & -- & -- & -- \\
    Dong72 & 266.566733 & 0.009 & -28.822761 & 0.02 & 0.014 $\pm$ 0.003 & -- & -- & -- & -- & -- & -- & -- & -- & 0.019 $\pm$ 0.004 & -- & -- & -- & -- \\
    qCG2 & 266.558149 & 0.01 & -28.827636 & 0.03 & 0.032 $\pm$ 0.006 & -- & -- & -- & -- & -- & -- & -- & -- & -- & -- & -- & -- & -- \\
    Hos21 & 266.551602 & 0.01 & -28.831895 & 0.02 & 0.017 $\pm$ 0.004 & -- & -- & -- & -- & -- & -- & -- & -- & -- & -- & -- & -- & -- \\
    Hos95 & 266.571180 & 0.02 & -28.839836 & 0.03 & 0.017 $\pm$ 0.004 & -- & -- & -- & -- & -- & -- & -- & -- & -- & -- & -- & -- & -- \\
  qF134 & 266.563493 & 0.009 & -28.834348 & 0.01 & 0.190 $\pm$ 0.015 & -- & 0.314 $\pm$ 0.023 & 0.293 $\pm$ 0.017 & 0.308 $\pm$ 0.018 & 0.385 $\pm$ 0.019 & 0.091 $\pm$ 0.016 & 0.106 $\pm$ 0.015 & 0.100 $\pm$ 0.017 & 0.101 $\pm$ 0.013 & 0.089 $\pm$ 0.029 & -- & -- & -- \\
  qF406 & 266.557711 & 0.02 & -28.814125 & 0.03 & 0.025 $\pm$ 0.005 & -- & -- & -- & -- & -- & -- & -- & -- & -- & -- & -- & -- & -- \\
  qF276 & 266.555909 & 0.02 & -28.824864 & 0.04 & 0.016 $\pm$ 0.005 & -- & -- & -- & -- & -- & -- & -- & -- & -- & -- & -- & -- & -- \\
    LHO029 & 266.561366 & 0.009 & -28.828725 & 0.02 & 0.019 $\pm$ 0.004 & -- & -- & -- & -- & -- & -- & -- & -- & 0.019 $\pm$ 0.005 & -- & -- & -- & -- \\
\hline \hline
\end{tabular}
}
\tablefoot{\footnotesize{
   All flux densities and their related uncertainties are in mJy units.
   Stellar ID from \citet{Clark2018_Quintuplet}, \citet{Dong2011}, \citet{Muno2009}, or \citet{Hosek2022}.
   }}
\end{center}
\end{sidewaystable*}

\setlength{\tabcolsep}{4pt}
\begin{sidewaystable*}
\footnotesize
\begin{center}
\caption{Spectral indices}
\label{Table_alphas}
\begin{tabular}{lcccccccccc}
\hline \hline
ID  & $\alpha^{04/10/16}_X$ &  $\alpha_X^{27/10/16}$ & $\alpha_X^{24/03/18}$ & $\alpha_C^{10/06/18}$  & $\alpha^{2018}$ & $\alpha_X^{20/05/22}$ & $\alpha_C^{03/06/22}$ & $\alpha_X^{02/07/22}$ & $\alpha_{CX}^{07/22}$ & $\alpha^{2022}$ \\
\hline
qF257 & $0.26 \pm 0.05$ & $0.06 \pm 0.11$ & $0.45 \pm 0.27$ & $0.09 \pm 0.09$ & -- & $0.16 \pm 0.06$ & $0.05 \pm 0.30$ & $0.48 \pm 0.18$ & $0.43 \pm 0.07$ & $0.43 \pm 0.23$ \\ 
qF241 & $0.69 \pm 0.03$ & $0.79 \pm 0.10$ & $0.59 \pm 0.08$ & $0.81 \pm 0.17$ & $0.82 \pm 0.16$ & $0.70 \pm 0.12$ & $0.68 \pm 0.31$ & $0.85 \pm 0.11$ & $0.60 \pm 0.05$ & $0.45 \pm 0.17$ \\ 
qF362 & $2.33 \pm 0.14$ & $1.78 \pm 0.17$ & $1.90 \pm 0.15$ & $2.39 \pm 0.56$ & $2.14 \pm 0.23$ & $0.64 \pm 0.12$ & $1.36 \pm 0.42$ & $0.89 \pm 0.22$ & $0.82 \pm 0.18$ & $1.00 \pm 0.23$ \\ 
qF240 & $1.08 \pm 0.48$ & $1.14 \pm 0.20$ & $1.25 \pm 0.30$ & $1.04 \pm 0.41$ & $0.91 \pm 0.23$ & $1.03 \pm 0.28$ & $0.44 \pm 0.61$ & $1.13 \pm 0.34$ & $0.72 \pm 0.11$ & $0.63 \pm 0.25$ \\ 
qF320 & $0.91 \pm 0.05$ & $-0.06 \pm 0.35$ & $0.77 \pm 0.25$ & $0.53 \pm 0.33$ & $0.59 \pm 0.24$ & $0.66 \pm 0.40$ & $1.03 \pm 0.30$ & $1.09 \pm 0.16$ & $0.64 \pm 0.10$ & $0.55 \pm 0.24$ \\ 
Muno7679 & $-0.14 \pm 0.27$ & $-0.06 \pm 0.43$ & $-0.37 \pm 0.35$ & $-0.64 \pm 0.28$ & $-0.28 \pm 0.19$ & $-0.51 \pm 0.28$ & $-0.48 \pm 0.34$ & $0.11 \pm 0.34$ & $-0.30 \pm 0.13$ & $-0.41 \pm 0.20$ \\ 
Dong66 & $0.00 \pm 0.52$ & $0.95 \pm 0.99$ & $-0.28 \pm 0.10$ & -- & $0.73 \pm 0.53$ & $1.07 \pm 0.18$ & -- & $1.32 \pm 0.47$ & $0.95 \pm 0.44$ & $-0.22 \pm 0.55$ \\ 
qF211 & $-0.02 \pm 0.56$ & $-0.01 \pm 0.58$ & $0.79 \pm 0.37$ & -- & $0.45 \pm 0.39$ & $-0.33 \pm 0.46$ & -- & $-0.09 \pm 1.20$ & $0.46 \pm 0.24$ & $0.25 \pm 0.42$ \\ 
qF256 & $1.24 \pm 0.74$ & $-0.09 \pm 0.23$ & $0.79 \pm 0.32$ & $0.53 \pm 0.21$ & -- & $0.43 \pm 0.46$ & -- & $0.45 \pm 0.99$ & $0.68 \pm 0.07$ & $0.69 \pm 0.43$ \\ 
qF231 & -- & $1.42 \pm 0.29$ & $0.58 \pm 1.30$ & -- & $0.49 \pm 0.49$ & $1.75 \pm 0.59$ & -- & $0.88 \pm 0.44$ & $0.41 \pm 0.34$ & $-0.13 \pm 0.46$ \\ 
qF270S & -- & $-0.23 \pm 0.52$ & $0.04 \pm 0.38$ & -- & -- & $1.78 \pm 0.73$ & -- & $1.59 \pm 1.20$ & $0.40 \pm 0.37$ & $1.27 \pm 0.65$ \\ 
qF235S & -- & $1.80 \pm 0.98$ & $-0.92 \pm 0.72$ & -- & $0.43 \pm 0.45$ & $-0.17 \pm 1.38$ & $-0.12 \pm 0.74$ & $1.15 \pm 0.61$ & $0.64 \pm 0.38$ & $0.42 \pm 0.50$ \\ 
qF307A & $0.70 \pm 0.43$ & $-0.74 \pm 0.91$ & $0.37 \pm 0.38$ & -- & $0.20 \pm 0.41$ & $0.60 \pm 0.51$ & $0.19 \pm 0.82$ & $0.72 \pm 1.26$ & $1.08 \pm 0.42$ & $0.53 \pm 0.50$ \\ 
qF344 & -- & -- & $1.15 \pm 0.58$ & -- & $0.03 \pm 0.52$ & $-0.43 \pm 0.66$ & $0.22 \pm 0.11$ & -- & $-0.79 \pm 0.13$ & $-0.93 \pm 0.38$ \\ 
qF235N & -- & $-1.77 \pm 1.01$ & $0.92 \pm 1.10$ & -- & $0.13 \pm 0.59$ & $1.51 \pm 0.35$ & -- & $0.93 \pm 0.61$ & $-0.05 \pm 0.36$ & $0.35 \pm 0.60$ \\ 
qF353E & -- & -- & $0.63 \pm 1.19$ & -- & -- & $2.17 \pm 0.76$ & -- & -- & -- & $-0.58 \pm 0.73$ \\ 
qF381 & -- & $0.02 \pm 0.96$ & -- & -- & -- & $1.09 \pm 0.65$ & -- & -- & -- & $-0.18 \pm 0.59$ \\ 
qCG1 & -- & -- & $2.09 \pm 1.35$ & $-0.61 \pm 0.46$ & $-1.14 \pm 0.46$ & $-0.16 \pm 0.74$ & -- & -- & $-0.90 \pm 0.68$ & $-1.45 \pm 0.54$ \\ 
qF278 & -- & -- & -- & -- & -- & -- & -- & -- & -- & $-0.23\pm0.58$ \\ 
qF309 & -- & $1.97 \pm 1.08$ & -- & -- & $0.30 \pm 0.63$ & $0.26 \pm 0.97$ & -- & -- & -- & $-0.13 \pm 0.76$ \\ 
LHO100 & -- & -- & -- & -- & -- & -- & -- & -- & -- & $0.00\pm0.66$ \\ 
qF274 & -- & $-0.60 \pm 0.57$ & -- & -- & $-0.28 \pm 0.66$ & $0.66 \pm 0.22$ & -- & $0.64 \pm 1.60$ & -- & -- \\ 
Hos332 & -- & -- & -- & -- & $-1.31 \pm 0.55$ & $-0.36 \pm 0.65$ & $-0.70 \pm 0.68$ & -- & $-0.47 \pm 0.68$ & $-0.79 \pm 0.62$ \\ 
LHO076 & -- & -- & -- & -- & -- & $0.47 \pm 0.10$ & -- & -- & -- & -- \\ 
WR102ca & -- & -- & -- & -- & -- & $1.79 \pm 0.14$ & -- & -- & -- & -- \\ 
LHO090 & -- & -- & -- & $-1.82 \pm 0.33$ & $-1.91 \pm 0.65$ & -- & $-0.81 \pm 0.09$ & $-1.87 \pm 0.51$ & -- & $-1.25 \pm 0.59$ \\ 
qF258 & -- & -- & $1.81 \pm 0.29$ & -- & -- & -- & -- & -- & -- & -- \\ 
qF134 & $1.91 \pm 0.45$ & $1.79 \pm 0.28$ & $1.81 \pm 0.31$ & -- & $2.87 \pm 0.65$ & $2.00 \pm 0.90$ & -- & $1.76 \pm 1.43$ & $0.90 \pm 0.48$ & -- \\ 
\hline \hline
\end{tabular}
\end{center}
\tablefoot{Values noted as $\alpha^{\mathrm{DD/MM/YY}}$ are obtained with the least squares linear fitting, and values in the columns noted as $\alpha^{\mathrm{year}}$ are obtained from Eqs. (3) and (4) from \citet{Arches_paper}.}
\end{sidewaystable*}

\end{appendix}

\end{document}